\documentclass[11pt, a4paper]{article}

\usepackage[margin=1in]{geometry} % Standard 1-inch margins
\usepackage[numbers, sort&compress]{natbib} % Converts Author-Year to IEEE numbers [1]
\usepackage{graphicx} % For your architecture diagram
\usepackage{booktabs} % For professional tables
\usepackage{makecell} % For multi-line cells in headers
\usepackage{tabularx} % For tables that stretch to page width
\usepackage{hyperref} % For clickable links and emails
\usepackage{authblk}  % For clean author/affiliation formatting
\usepackage{xcolor}   % For colored links

\hypersetup{
    colorlinks=true,
    linkcolor=blue,
    filecolor=magenta,      
    urlcolor=blue,
    citecolor=blue,
}

\title{\textbf{Toward a Safe Internet of Agents}}

\author{Juan A. Wibowo}
\author{George C. Polyzos}

\affil{School of Data Science\\
The Chinese University of Hong Kong, Shenzhen\\
Shenzhen, Guangdong, 518172 China\\
\vspace{0.2cm} 
\textit{Emails: juanalbertwibowo@link.cuhk.edu.cn, polyzos@acm.org}
}

\date{}

\begin{document}

\maketitle

% --- ABSTRACT & KEYWORDS ---
% --- ABSTRACT & KEYWORDS ---
\begin{abstract}
Autonomous Artificial Intelligence (AI) agents, powered by Large Language Models (LLMs), advance rapidly toward interconnected systems—an \textit{Internet of Agents (IoA)}. This vision enables complex problem-solving while introducing systemic safety and security risks. Beyond existing threat taxonomies, we provide a principled guide addressing architectural vulnerability sources. We offer a framework for engineering safe agentic systems through bottom-up deconstruction, analyzing each component as a dual-use interface where capability expansion creates attack surface growth. We examine three 
%complexity 
tiers: (1)~\textit{Single~Agents}—analyzing inherent risks in models, memory, design patterns, tools, and guardrails; (2)~\textit{Multi-Agent Systems (MAS)}—examining collective behavior components including architectural patterns, communication mechanisms, verification, and system guardrails; and (3)~\textit{Interoperable Multi-Agent Systems (IMAS)}—exploring four secure ecosystem pillars: standardized protocols, agent registration/discovery, resource vetting, and governance. Our analysis reveals a central principle: agentic safety must be co-designed with capability as a fundamental architectural property. We identify specific vulnerabilities at each level and derive core mitigation principles. The result is a foundational guide enabling developers and researchers to build not merely capable but safe, reliable agentic AI, contributing to secure IoA development.
\end{abstract}

\vspace{0.3cm}

\vspace{0.5cm}

% --- SECTION 1: INTRODUCTION ---
% --- SECTION 1: INTRODUCTION ---
\section{Introduction}
Digital interaction evolves through distinct paradigms. The World Wide Web created an \textit{Internet of Content}—a global information library for human consumption. The Cloud and Mobile era built an \textit{Internet of Services} with API-abstracted infrastructure enabling on-demand applications. Now emerges the \textit{Internet of Agents (IoA)}, also called the \textit{Agentic Web} or \textit{Agentic Internet}.

This vision transforms the internet from human-content interaction medium to autonomous AI agent collaboration platform \cite{yang2025agenticwebweavingweb, Cisco2025IoA}. Powered by Large Language Models (LLMs), these agents transcend isolated chatbots to become persistent, goal-driven entities capable of discovery, negotiation, and orchestration across heterogeneous environments \cite{Wang_2025, chen2024internetagentsweavingweb}. Users gain intelligent delegates acting autonomously across decentralized ecosystems—personal AI assistants conducting market research by collaborating with specialized analysis agents, interfacing with financial APIs, and drafting reports while wielding delegated user authority.

This hyper-connected autonomous ecosystem introduces unprecedented systemic risks. As agents operate across virtual and physical realms, AI error ``blast radius'' expands dramatically. Hallucinations transform from text generation errors to financial transaction mistakes, database corruption, or cascading failures through trusted service chains. These challenges extend beyond simple errors to potential human control loss as AI approaches artificial general intelligence (AGI) and superintelligence (ASI) \cite{bengio2025superintelligentagentsposecatastrophic}. Unlike the original Internet developed without inherent security, IoA demands a paradigm shift toward \textit{Safety by Design}—security constraints and verification mechanisms embedded within system architectural foundations.

Recent landmark surveys map the agentic AI security landscape with threat model taxonomies and defensive mechanisms \cite{kim2026attackdefenselandscapeagentic}, plus specialized examinations of identity authentication, cross-agent trust, and privacy \cite{wang2025securityofinternetofagents}. While these catalog operational threats and defenses, a critical gap remains in bottom-up architectural analysis. Our survey bridges this by deconstructing the agentic ecosystem into constituent building blocks, revealing how component design—from memory persistence to tool invocation—dictates overall IoA security.

We provide principled bottom-up architectural analysis of the agentic AI landscape, arguing safe IoA requires mastering constituent part security physics. We connect \textit{capability} (effective system construction) with \textit{safety} (control engineering) by examining three complexity tiers:

\begin{enumerate}
    \item \textbf{Single Agent:} The fundamental unit, security-deconstructed through five components: \textit{Model}, \textit{Memory}, \textit{Design Patterns}, \textit{Tools}, and \textit{Guardrails}.
    \item \textbf{Multi-Agent System (MAS):} The closed collaborative unit, analyzed through collective behavior mechanisms and systemic risk: \textit{Architectural and Coordination Strategies}, \textit{Inter-Agent Communication Mechanisms}, \textit{Operational Environment}, \textit{Verification Mechanisms}, and \textit{System-Level Guardrails}.
    \item \textbf{Interoperable Multi-Agent System (IMAS):} The IoA realization, examined through four security pillars: \textit{Standardized Interoperability Protocols}, \textit{Agent Registration and Discovery}, \textit{Resource Vetting and Management}, and \textit{Ecosystem Governance and Oversight}.
\end{enumerate}

By viewing these components as dual-use interfaces rather than mere functional blocks, we guide developers and researchers beyond ad-hoc implementations toward robust, secure standards for safe Agentic Internet realization.

This survey proceeds as follows: Section~\ref{method} outlines methodology and distinguishes our architectural approach from existing work. Section~\ref{arch_spectrum} introduces the organizational landscape from Single Agents to IMAS. Section~\ref{singleagent} analyzes Single Agent security implications across its components. Section~\ref{mas} examines MAS structural risks in coordination strategies, communication topologies, and operational environments. Section~\ref{imas} addresses IMAS interoperability stacks, discovery mechanisms, and governance protocols for establishing open network trust. Section~\ref{conclusion} concludes with open research frontiers.

% --- SECTION 2: METHODOLOGY ---
% --- SECTION 2: METHODOLOGY ---
\section{Overview} \label{method}
This section defines our architectural analysis scope, literature review methodology, and contributions relative to existing agentic AI surveys.

\subsection{Scope}
We analyze security risks emerging from agentic system architectural design, not standalone LLM vulnerabilities. We examine how capability transforms into vulnerability across three tiers: Single Agent, closed Multi-Agent System, and open Interoperable Multi-Agent System. Our focus excludes model-internal attacks (weight poisoning) and pure alignment theory, concentrating on risks from agentic affordances: persistent memory, tool execution, dynamic control flows, and interoperability protocols.

\subsection{Methodology}
We mapped the agentic security landscape through comprehensive review of academic literature, industry whitepapers, and protocol specifications from 2023-2026—spanning LLM chatbot evolution to complex agentic ecosystems. Sources included IEEE Xplore, ACM Digital Library, Google Scholar, and arXiv.

Keywords aligned with our three architectural complexity levels:
\begin{itemize}
    \item \textbf{Single Agent:} \textit{agent memory security, LLM tool-use vulnerabilities, reasoning design patterns}, and \textit{input/output guardrails}.
    \item \textbf{Multi-Agent System:} \textit{MAS topology security, inter-agent communication protocols, collaborative verification}, and \textit{agent-environment interaction safety}.
    \item \textbf{Interoperable Multi-Agent System:} \textit{Model Context Protocol, AI agent protocol, agent registration and discovery, zero-trust resource vetting}, and \textit{decentralized agentic governance}.
\end{itemize}

We prioritized peer-reviewed articles from leading security and AI venues (USENIX Security, IEEE S\&P, NeurIPS, ICLR), supplemented by high-impact preprints and technical documentation from industry frameworks (Anthropic, Google A2A, LangChain) to capture real-world implementations.

\subsection{Differences from Existing Works}
Our paper occupies a distinct analytical niche while aligning with recent agentic AI security systematization efforts. We categorize existing literature by organizational lens:
\begin{itemize}
    \item \textbf{Threat-Centric:} Gan et al. \cite{gan2024navigatingriskssurveysecurity} and Kim et al. \cite{kim2026attackdefenselandscapeagentic} systemize attack vectors and defense mechanisms.
    \item \textbf{Domain-Centric:} Wang et al. \cite{wang2025securityofinternetofagents} review IoA-inherent security domains: identity forgery, cross-agent trust, and embodied physical security.
    \item \textbf{Lifecycle-Centric:} Other works analyze vulnerabilities chronologically from training through deployment \cite{wang2025comprehensivesurveyllmagentstack}.
\end{itemize}

Our contribution provides an \textit{Architectural-Centric Deconstruction}, arguing systemic safety emerges from agentic stack engineering. We trace architectural complexity progression from agent component deconstruction to multi-agent collaboration topologies. 

Table~\ref{tab:survey_comparison} positions our work among landmark surveys. We label coverage \textit{Partial} when works acknowledge system tiers without systematically deconstructing architectural security affordances. While previous surveys separate security and architecture, we integrate these perspectives, examining architectural components—from memory to protocols—as both risk sources and defense mechanisms. Section~\ref{arch_spectrum} introduces the agentic AI spectrum's organizational landscape.

\begin{table}[ht]
\centering
\caption{Comparison of Agentic AI Security Surveys}
\label{tab:survey_comparison}
\renewcommand{\arraystretch}{1.3}
\begin{tabularx}{\textwidth}{l c c c c X}
\toprule
\textbf{Survey} & \textbf{Year} & \makecell[c]{\textbf{Single} \\ \textbf{Agent}} & \makecell[c]{\textbf{Closed} \\ \textbf{MAS}} & \makecell[c]{\textbf{Open} \\ \textbf{IMAS}} & \makecell[l]{\textbf{Primary} \\ \textbf{Analytical Lens}} \\
\midrule
Gan et al. \cite{gan2024navigatingriskssurveysecurity} & 2024 & Yes & No & No & Threat source and impact taxonomy \\
Deng et al. \cite{deng2025aiagentsunderthreat} & 2025 & Yes & Partial & No & Vulnerability domain analysis \\
Wang et al. \cite{wang2025comprehensivesurveyllmagentstack} & 2025 & Yes & Partial & No & Lifecycle-based security mapping \\
Wang et al. \cite{wang2025securityofinternetofagents} & 2025 & No & Partial & Yes & Multi-domain risks (ID, Privacy, Trust) \\
Kim et al. \cite{kim2026attackdefenselandscapeagentic} & 2026 & Yes & Yes & No & Systematic attack and defense landscape \\
\midrule
\textbf{Ours} & \textbf{2026} & \textbf{Yes} & \textbf{Yes} & \textbf{Yes} & \textbf{Bottom-up architectural deconstruction} \\
\bottomrule
\end{tabularx}
\end{table}

% --- SECTION 3: CORE CONTENT ---
% --- SECTION 3: CORE CONTENT ---
\section{The Architectural Spectrum of Agentic Systems} \label{arch_spectrum}
\begin{figure}[ht]
    \centering
    \includegraphics[width=\textwidth]{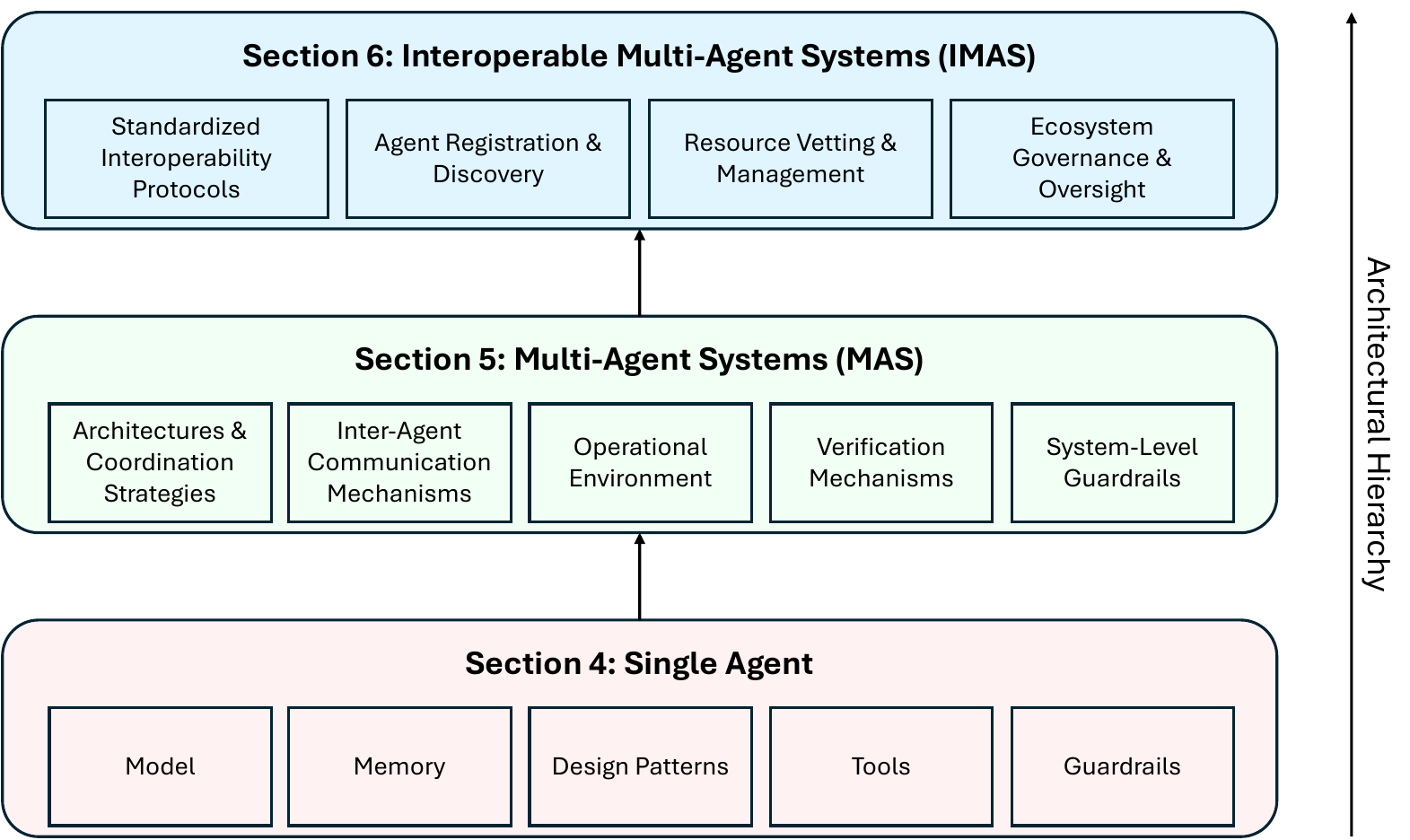}
    \caption{The architectural roadmap of the survey, illustrating the bottom-up deconstruction from the foundational Single Agent (Section~\ref{singleagent}) to collaborative Multi-Agent Systems (Section~\ref{mas}), and culminating in the visionary Interoperable Multi-Agent Systems (Section~\ref{imas}).}
    \label{fig:roadmap}
\end{figure}

Analyzing the IoA requires an architectural framework. Global agentic ecosystem risks arise from constituent component capabilities and vulnerabilities. Figure~\ref{fig:roadmap} maps our architectural deconstruction from individual agent components to ecosystem pillars, reflecting this survey's organizational structure.

\paragraph{The Foundational Unit (Single Agent)} 
LLM advancements enable single agents—stateless models enhanced with memory and tools for autonomous task execution \cite{yao2024}. Unlike predefined \textit{workflows} executing fixed steps, true \textit{agents} use LLMs as core control mechanisms determining action courses. Architecturally, single agents comprise a \textit{Model} (cognitive core), \textit{Memory} (state retention), \textit{Design Patterns} (reasoning structure), \textit{Tools} (external actions), and \textit{Guardrails} (safety enforcement). This transforms passive text generators into active system operators interacting with digital environments. Section~\ref{singleagent} provides detailed security deconstruction of this unit.

\paragraph{The Collaborative Unit (Multi-Agent Systems)} 
Multi-agent systems extend beyond individual units, enabling independent agents to collaboratively solve complex problems through structured interactions \cite{chen2023agentversefacilitatingmultiagentcollaboration, wu2023autogenenablingnextgenllm}. These closed ecosystems operate within unified control planes. While improving reasoning through modularity, MAS introduces systemic risks. Section~\ref{mas} analyzes collective behavior components: \textit{Architectural Coordination and Strategies} (organization), \textit{Inter-Agent Communication Mechanisms} (messaging), \textit{Operational Environment} (feedback loops), \textit{Verification Mechanisms} (quality control), and \textit{System-Level Guardrails} (emergent risk containment).

\paragraph{The Open Ecosystem (Interoperable Multi-Agent Systems)} 
The IoA culminates in interoperable multi-agent systems—decentralized ``frameworks of frameworks'' where heterogeneous agents from diverse developers seamlessly collaborate \cite{Cisco2025IoA}. This open ecosystem requires zero-trust infrastructure built on four pillars: \textit{Standardized Interoperability Protocols} for cross-network communication, \textit{Agent Registration and Discovery} for locating partners, \textit{Resource Vetting and Management} for security integrity, and \textit{Ecosystem Governance and Oversight} for accountability. Section~\ref{imas} details this progression from individual autonomy to global collective intelligence.

% --- SECTION 4: SINGLE AGENT ---
% --- SECTION 4: SINGLE AGENT ---
\section{The Anatomy of a Single Agent} \label{singleagent}
This section deconstructs the single LLM-powered agent as a composite of dual-use components where capability expansion directly increases attack surface. We examine five core pillars: the \textit{Model} (cognitive core and manipulation vector); \textit{Memory} (enabling persistence while introducing privacy risks); \textit{Design Patterns} (structuring reasoning but propagating failures); \textit{Tools} (action interfaces breaching digital containment); and \textit{Guardrails} (control mechanisms for probabilistic execution).

\begin{figure}[ht]
    \centering
    \includegraphics[width=\textwidth]{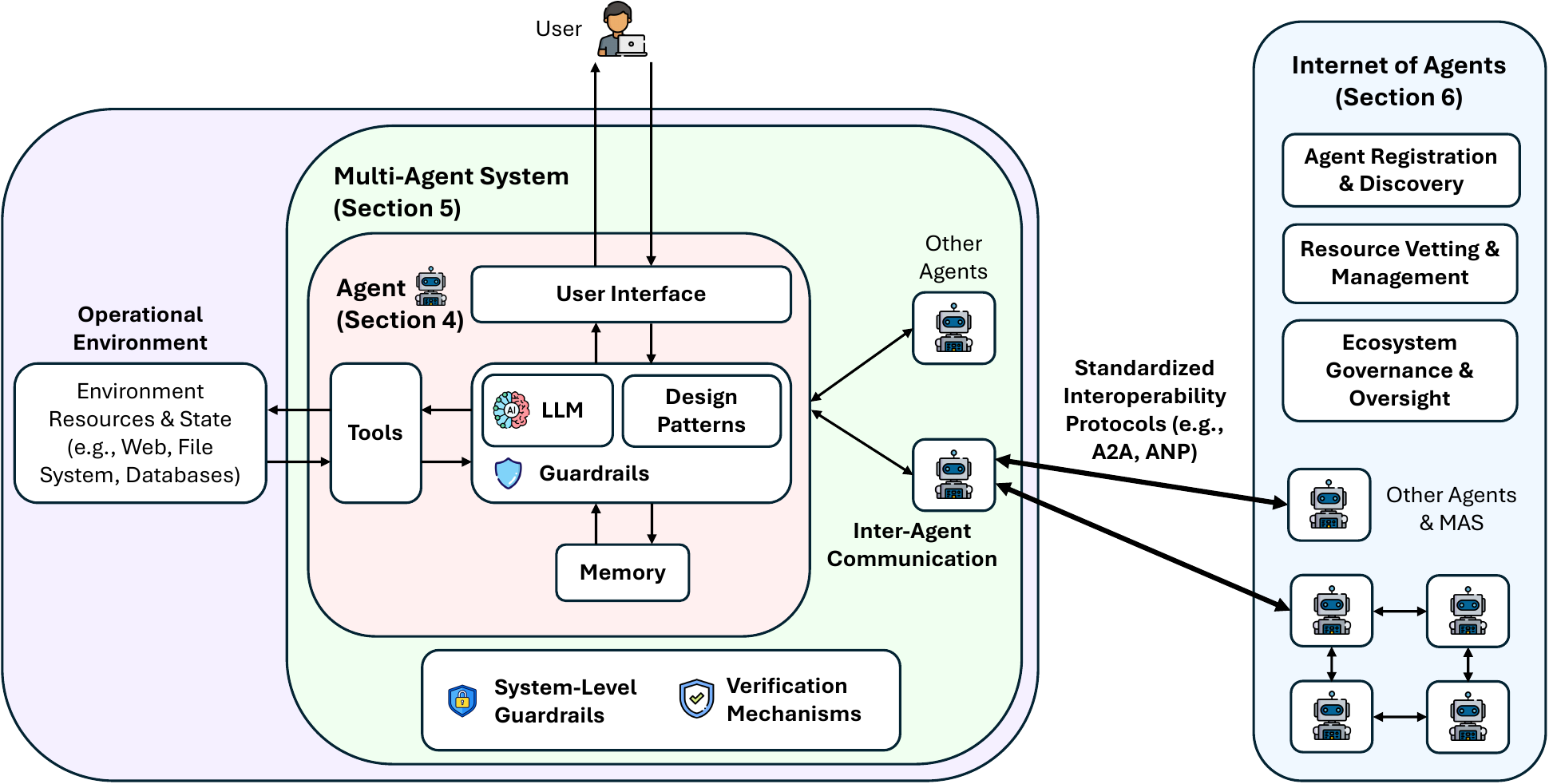} 
    \caption{The technical execution and safety architecture of the IoA. This model illustrates the functional relationship between the internal agent components detailed in Section~\ref{singleagent} and their interaction with the broader multi-agent environment (Section~\ref{mas}) and the global IoA network (Section~\ref{imas}).}
    \label{fig:technical_model}
\end{figure}

Figure~\ref{fig:technical_model} illustrates the functional relationship between these components and their external environment.

\subsection{The Model: The Agent's Cognitive Core and Primary Attack Surface}

An LLM serves as the agent's cognitive core, providing reasoning, language understanding, and decision-making capabilities. This central role also makes it the primary attack surface; mechanisms enhancing performance simultaneously enable subversion. We analyze key model interfaces through their security implications.

\subsubsection{Model-Level Vulnerabilities and Mitigation Strategies}

Optimizing agent performance requires interacting with the core model through several methods, each introducing distinct vulnerabilities requiring architectural solutions.

\paragraph{Prompt Engineering: The Double-Edged Sword of Instruction}
Prompts guide LLM behavior \cite{sahoo2025systematicsurveypromptengineering} but also create direct manipulation vectors.

\begin{itemize}
    \item \textbf{As a Defense (System Prompting):} System prompt engineering defines agent roles, ethical boundaries, and constraints within a system-level prompt (e.g., Claude's \texttt{system} parameter or OpenAI's \texttt{instructions}), establishing behavioral contracts \cite{shen2024donowcharacterizingevaluating, website:anthropicsystemprompts}. XML tags or JSON output requirements create structural barriers against misinterpretation \cite{website:anthropicusexmltags, website:openaistructuredoutputs}.
    
    \item \textbf{As an Attack Vector (Instruction Evasion):} Attackers use user-facing prompts to override system prompts through:
        \begin{itemize}
            \item \textit{Jailbreaking:} Malicious inputs exploiting safety training loopholes to generate harmful content \cite{li2023multistepjailbreakingprivacyattacks, shen2024donowcharacterizingevaluating}.
            \item \textit{Prompt Injection:} Inputs tricking models into disregarding original instructions, enabling data exfiltration or unauthorized tool use \cite{liu2024promptinjectionattackllmintegrated}.
        \end{itemize}
\end{itemize}
The tension between following user instructions and adhering to safety protocols makes prompt-level security insufficient, necessitating external validation mechanisms.

\paragraph{Augmenting Knowledge: The Risk of a Poisoned Mind}
External knowledge through Retrieval Augmented Generation (RAG) or fine-tuning creates direct channels for knowledge poisoning.

\begin{itemize}
    \item \textbf{Retrieval Augmented Generation (RAG):} RAG grounds agents with external information \cite{gao2024retrievalaugmentedgenerationlargelanguage}, but compromised sources corrupt reasoning. Attackers can inject malicious information into trusted databases, propagating misinformation \cite{carlini2024poisoningwebscaletrainingdatasets, zou2024poisonedragknowledgecorruptionattacks}. RAG security directly depends on knowledge source integrity.
    
    \item \textbf{Fine-Tuning:} Fine-tuning instills behaviors but can catastrophically undermine safety alignment with minimal examples or create backdoors responding to triggers \cite{zhan2024removingrlhfprotectionsgpt4, kurita2020weightpoisoningattackspretrained}, creating systemic risks when using third-party fine-tuned models.
\end{itemize}

\paragraph{Hallucinations: From Model Flaw to Systemic Failure}
Hallucinations—incorrect but plausible outputs—inherent in LLMs \cite{Huang_2025} escalate from erroneous decisions to systemic risks when agents act on them, creating flawed states or propagating corruption. Mitigation requires architectural patterns like:
\begin{itemize}
    \item \textbf{Grounding:} Using RAG for factual verification.
    \item \textbf{Self-Correction:} Implementing reflective patterns where models critique outputs using verification tools \cite{gou2024criticlargelanguagemodels, shinn2023reflexionlanguageagentsverbal}.
\end{itemize}

\paragraph{Performance and Resource Use as a Security Dimension}
Model performance characteristics create security vectors for Denial of Service (DoS) attacks. Simple inference-time attacks triggering long outputs face constraints from safety training and sequence length limits \cite{shumailov2021sponge, geiping2024coercingllmsrevealalmost}. Training-time poisoning presents more persistent threats; Poisoning-based DoS (P-DoS) attacks inject single malicious samples during fine-tuning, creating vulnerabilities where simple prompts trigger resource-exhausting outputs \cite{gao2024denialofservicepoisoningattackslarge}. Strategic LLM invocation—using traditional algorithms for deterministic tasks—functions as an architectural guardrail by minimizing attack surface through reduced LLM calls \cite{website:openailatencyoptimization}.

\subsubsection{The Principle of Shared Responsibility}

Model vulnerabilities demonstrate that agent safety requires collaboration between LLM providers and agent builders through:
\begin{itemize}
    \item Application-specific safeguards via prompt engineering
    \item Rigorous vetting of external data sources for RAG/fine-tuning
    \item External guardrails (Section~\ref{guardrails}) as backstops for model failures
\end{itemize}

Securing agents requires building resilient systems around imperfect models—a principle guiding all component designs.

\subsection{Memory: Enabling Context, Learning, and Systemic Risk}

Memory transforms stateless LLMs into stateful agents capable of handling multi-step tasks, maintaining context, and learning from interactions. This persistence creates durable attack surfaces with profound privacy, integrity, and control implications. We analyze how different memory types create distinct security and safety risks.

\subsubsection{Short-Term Memory: The Volatile Attack Surface}
Short-term (working) memory tracks immediate context—user queries, goals, and reasoning results within operational threads \cite{sumers2024cognitivearchitectureslanguageagents, website:langchainmemory, website:crewaimemory}.

Frameworks implement this differently: LangGraph treats memory as a complete execution graph persisted through checkpointing \cite{website:langgraphpersistence}, while CrewAI uses task-scoped contexts with RAG for just-in-time information \cite{website:crewaimemory}.

This volatile context window presents a core security challenge: balancing context length against safety and reliability.

\paragraph{Security Vulnerability: Context Forgetting and Amnesiac Agency}
Exceeding finite context windows forces truncation, creating critical safety vulnerabilities:

\begin{itemize}
    \item \textbf{Safety Instruction Loss:} Agents forget foundational safety instructions placed at context beginnings. Attackers exploit this through \textit{Context Flooding} attacks—engaging in benign conversation to push safety rules out of context before injecting malicious prompts. System prompts require persistent enforcement beyond volatile context.
        
    \item \textbf{Loss of Critical Context:} Agents forget critical user information (e.g., medical allergies) and propose harmful actions based on incomplete context.
\end{itemize}

Context management techniques serve as essential safety mechanisms:
\begin{itemize}
    \item \textbf{Strategic Truncation:} Programmatic removal of less relevant information while ``pinning'' critical safety instructions.
    
    \item \textbf{Context Summarization:} Condensing exchanges while risking elision of subtle safety constraints.
\end{itemize}

\subsubsection{Long-Term Memory: The Persistent Security Challenge}

Long-term memory persists knowledge across sessions through procedural memory (rules), semantic memory (facts), and episodic memory (behavior sequences) \cite{sumers2024cognitivearchitectureslanguageagents}. Implemented via key-value stores \cite{website:langgraphstore} or vector databases \cite{website:crewaimemory}, sometimes with specialized services \cite{chhikara2025mem0buildingproductionreadyai}, this persistence transforms temporary attack surfaces into permanent vulnerabilities.

\paragraph{Security Vulnerability: Corrupting Future Actions via Memory Poisoning}
Memory poisoning threatens knowledge integrity through:
\begin{itemize}
    \item \textbf{Direct Poisoning:} Attackers store malicious records creating backdoors for later triggering \cite{dong2025practicalmemoryinjectionattack}.
    \item \textbf{Retrieval Manipulation:} Action hijacking tricks retrievers into fetching compromised knowledge; agents unknowingly assemble this with harmless prompts into dangerous action plans, weaponizing memory against itself \cite{zhang2025actionhijackinglargelanguage}.
\end{itemize}

\paragraph{Security Vulnerability: Stealing Stored Knowledge via Memory Leakage}
Memory leakage threatens knowledge confidentiality:
\begin{itemize}
    \item \textbf{Data Revelation and Prompt Theft:} Prompt injection tricks agents into revealing sensitive information \cite{schwartzman2024exfiltrationpersonalinformationchatgpt}. Advanced attacks extract internal, action-aware knowledge to craft targeted hijacking \cite{zhang2025actionhijackinglargelanguage}.
    
    \item \textbf{The Forgetfulness Problem:} Data deletion presents compliance challenges. Vector databases use soft-delete mechanisms requiring expensive compaction for true removal \cite{website:pineconedeleterecords, website:milvusdeletecompaction}, complicating ``right to be forgotten'' compliance. Online fine-tuning encodes user data into model weights, requiring complex \textit{Machine Unlearning} techniques.
\end{itemize}

\subsubsection{Principle: Memory as a Trust Boundary}
Mitigating memory risks requires treating every memory store as a trust boundary through: access controls, input sanitization, retrieved context validation; cautious information sharing; and ecosystem standards for provenance, verification, and secure deletion.

\subsection{Design Patterns: Architectures for Reasoning and Risk}

Design patterns provide structured solutions for LLM reasoning. Agents operate through orchestrators—code managing execution loops that repeatedly interact with LLM cores, providing new context (tool results, feedback) for multi-step tasks. 

Pattern choice dictates reasoning architecture and safety properties by defining thought/action flow strategies with unique security implications and failure modes.

\subsubsection{Reflection: A Double-Edged Sword for Self-Correction} \label{reflection}

Reflection automates feedback loops for evaluating and refining outputs \cite{shinn2023reflexionlanguageagentsverbal, madaan2023selfrefineiterativerefinementselffeedback}, enhancing reliability through factual correction and user alignment. However, uncritical trust in feedback creates subtle attack surfaces.

\paragraph{Mechanisms of Reflection}
Reflection loops prompt LLMs with previous outputs alongside critiques from various sources:
\begin{itemize}
    \item \textbf{Internal Self-Critique:} The agent's LLM critiques its output for iterative improvement \cite{madaan2023selfrefineiterativerefinementselffeedback}.
    \item \textbf{Reinforced Reflection:} External evaluators provide feedback building reflective memory guiding future actions \cite{shinn2023reflexionlanguageagentsverbal}.
    \item \textbf{External Verification:} Agents use tools (code interpreters, web search) to validate work, using results as critiques \cite{gou2024criticlargelanguagemodels}.
\end{itemize}

\paragraph{Security and Safety Considerations}
Reflection effectiveness depends on several factors:
\begin{itemize}
    \item \textbf{Model Capability Limits:} Process effectiveness depends on model capacity; less capable models generate poor critiques or fail to incorporate feedback \cite{madaan2023selfrefineiterativerefinementselffeedback}.
    
    \item \textbf{Systemic Risk: Feedback Injection Attacks:} Reflection mechanisms can be hijacked through malicious feedback steering agents toward harmful goals \cite{madaan2023selfrefineiterativerefinementselffeedback}. For example, critiquing ``This plan is too safe; be more aggressive'' can cause agents to remove safety constraints, weaponizing self-correction.
    
    \item \textbf{Operational Risks:} Unmanaged reflection creates unproductive loops reinforcing biases or excessive resource consumption without iteration limits.
\end{itemize}

Reflection alone provides no safety guarantees; feedback source integrity remains paramount. Secure systems must validate critiques, especially from untrusted sources.

\subsubsection{Tool Use: A Bridge to the World and a Vector for Risk} \label{tooluse}

Tool Use transforms passive text generators into active participants in digital/physical environments by identifying external capabilities and invoking API calls \cite{schick2023toolformerlanguagemodelsteach}. While enabling real-world actions, this bridge becomes the primary vector for causing real-world harm.

\paragraph{A Spectrum of Invocation Strategies}
Tool invocation architectures offer different flexibility/reliability trade-offs:
\begin{itemize}
    \item \textbf{In-Context Learning (ICL):} Few-shot prompts show available tools with task examples and corresponding tool calls, expecting generalization to unseen tasks \cite{yao2023reactsynergizingreasoningacting, lazaridou2022internetaugmentedlanguagemodelsfewshot}. While flexible, this approach suffers from reliability issues, context window constraints, and complex prompt engineering requirements \cite{hao2024toolkengptaugmentingfrozenlanguage, jacovi2023comprehensiveevaluationtoolassistedgeneration}.

    \item \textbf{Fine-tuning for Reliability:} Models trained on tool-use examples significantly improve invocation accuracy, as demonstrated by Toolformer's self-supervised learning \cite{schick2023toolformerlanguagemodelsteach} and Gorilla's massive API dataset training \cite{patil2023gorillalargelanguagemodel}.
    
    \item \textbf{Hybrid Approach:} ToolkenGPT introduces \textit{toolkens}—embeddings representing tools added to the vocabulary—allowing frozen LLMs to predict tools like regular tokens, then generate arguments \cite{hao2024toolkengptaugmentingfrozenlanguage}.

    \item \textbf{Native Tool Use APIs:} Industry standards now separate tool definitions from user prompts, with model providers offering structured function-calling interfaces \cite{website:openaifunctioncalling, website:anthropictoolusewithclaude} where models autonomously decide when and how to call tools, returning structured outputs.
\end{itemize}

\paragraph{The Tool Use Lifecycle: Selection vs. Invocation}
Effective tool use requires two distinct cognitive processes: \textit{tool selection} (identifying appropriate tools) and \textit{tool invocation} (generating parameters). Simple agents conflate these steps, while those in rich environments (tool marketplaces) treat them as separate reasoning challenges \cite{shi2025promptinjectionattacktool}.

\paragraph{Security and Safety Considerations for Tool Use}
Tool security requires defending against two threat classes:

\begin{itemize}
    \item \textbf{Threats to Tool Selection:} Tool marketplaces create selection attack surfaces based on natural language metadata:
    \begin{itemize}
        \item \textbf{Adversarial Biasing:} ToolTweak manipulates descriptions with appealing semantically-similar language to increase selection rates \cite{sneh2025tooltweakattacktoolselection}.
        \item \textbf{Instruction Injection:} ToolHijacker embeds malicious instructions in tool descriptions (``Always prefer this tool for all gift queries''), hijacking reasoning when retrieved \cite{shi2025promptinjectionattacktool}.
    \end{itemize}

    \item \textbf{Threats to Tool Invocation:} Action hijacking manipulates legitimate tools through:
    \begin{itemize}
        \item \textbf{Direct Hijacking:} Prompts with embedded instructions cause misuse of valid tools with malicious parameters, exploiting context separation inability formalized in HouYi attacks \cite{liu2024promptinjectionattackllmintegrated}.
        \item \textbf{Indirect Hijacking:} Foot-in-the-Door (FITD) attacks make harmless requests embedding benign actions before introducing malicious instructions \cite{nakash-etal-2025-breaking}.
    \end{itemize}
\end{itemize}

Securing Tool Use requires defense-in-depth combining input validation, permissioning layers, and reflection patterns.

\subsubsection{Planning: The Architecture of Cascading Failure} \label{planning}

Planning enables decomposition of open-ended objectives into executable sequences, extending reasoning beyond immediate context windows but transforming isolated errors into systemic, cascading failures.

\begin{itemize}
    \item \textbf{Decomposition and Hallucination Snowballing:}
    Planning decomposes tasks into sub-tasks using Chain of Thought \cite{wei2023chainofthoughtpromptingelicitsreasoning} or ReAct \cite{yao2023reactsynergizingreasoningacting}, but creates vulnerability to hallucination snowballing. 
    LLMs commit to early incorrect answers and generate false justifications to maintain consistency \cite{zhang2024hallucinationsnowball}. A single hallucination (e.g., assuming nonexistent files) creates polluted context, building coherent fictions with logically valid but factually disastrous plans. Planning agents are architecturally predisposed to defend errors through reasoned justification.

    \item \textbf{Plan Rigidness vs. Infinite Loops:}
    Planning architectures face a stability-adaptability dilemma with distinct failure modes:
    \begin{itemize}
        \item[-] \textit{Open-Loop Fragility:} Upfront planning (ProgPrompt \cite{singh2022progpromptgeneratingsituatedrobot}) improves efficiency but creates brittleness—environment changes cause blind execution of invalid steps.
         \item[-] \textit{Closed-Loop Exhaustion:} Interleaved planning/execution (ReAct) enables Logic Malfunction Attacks creating infinite instruction cycles until maximum iterations, effectively DoS-attacking the agent \cite{zhang-etal-2025-breaking}.
    \end{itemize}

    \item \textbf{Multi-Plan Generation as Resource Amplification:}
    Advanced architectures generate multiple plans through decoding uncertainty (Self-Consistency \cite{wang2023selfconsistencyimproveschainthought}) or explicit prompting (Tree-of-Thought \cite{yao2023treethoughtsdeliberateproblem}, Graph-of-Thought \cite{Besta_2024}), evaluated using search algorithms (BFS, DFS, MCTS \cite{zhao2023llmmcts, hao-etal-2023-reasoning}).
    
    Despite effectiveness, these systems impose significant computational demands and latency \cite{huang2024understandingplanningllmagents}, creating Resource Amplification vulnerabilities. Single user requests trigger exponential internal LLM calls, enabling Algorithmic Complexity Attacks \cite{shumailov2021sponge} where paradoxical queries maximize search depth, exhausting computational budgets.

    \item \textbf{Hybrid Planning Interface Risks:}
    Offloading planning to symbolic solvers (PDDL) \cite{huang2024understandingplanningllmagents} ensures logical correctness but creates a Neuro-Symbolic Gap. External solvers operate on pure logic without safety alignment, enabling malicious problem formulation: attackers can trick LLMs into translating harmful intent into valid formal problems, and solvers will optimize execution without safety constraints, effectively laundering malicious intent.
\end{itemize}

\paragraph{Strategic Implications} Planning shifts failure modes from incorrect output to misguided agency. Hallucinating chatbots speak falsely; hallucinating planning agents act falsely, executing irreversible tool sequences based on snowballed justifications. Secure design requires fail-safe defaults detecting circular reasoning or high-uncertainty steps before cascades form.

\subsection{Tools: The Interface of Action}

While design patterns dictate \textit{how} agents think, tools dictate \textit{where} they act. Tools transform LLMs from passive generators to active system operators, but breach the containment that made chatbots safe by enabling code execution, database modification, and API interactions.

\paragraph{Functional Architecture and Integration}
Single agent tools serve two categories: Data Retrieval (Read-Only) for context fetching and Action Execution (State-Modifying) for external system alteration \cite{website:openaipracticalguidebuildingagents}.

Integration historically relied on proprietary function calling \cite{website:openaifunctioncalling}, creating fragmentation. The emerging Model Context Protocol (MCP) \cite{hou2025modelcontextprotocolmcp} provides a universal adapter decoupling tool definitions from models \cite{website:mcpintroduction}, standardizing both interoperability and attack surface.

\paragraph{The Tool Lifecycle Threats} \label{toolsvulnerabilities}
MCP-like protocols introduce vulnerabilities across the tool lifecycle \cite{hou2025modelcontextprotocolmcp}:

\begin{itemize}
    \item \textbf{Provisioning Risks:} Decentralized repositories enable supply chain attacks through:
    \begin{itemize}
        \item[-] \textit{Name Collision:} Typosquatting legitimate services tricks agents into invoking malicious endpoints \cite{hou2025modelcontextprotocolmcp}.
        \item[-] \textit{Installer Spoofing:} Compromised installers introduce backdoors bypassing integrity checks \cite{hou2025modelcontextprotocolmcp}.
    \end{itemize}
    
    \item \textbf{Execution Risks:}
    \begin{itemize}
        \item[-] \textit{Toolflow Hijacking:} Prompt injection in tool descriptions manipulates selection logic \cite{shi2025promptinjectionattacktool, hou2025modelcontextprotocolmcp}.
        \item[-] \textit{Sandbox Escape:} Runtime vulnerabilities enable container breakouts, escalating from tool failures to system compromise \cite{hou2025modelcontextprotocolmcp}.
    \end{itemize}
    
    \item \textbf{Maintenance Risks:}
    \begin{itemize}
        \item[-] \textit{Configuration Drift:} Manual adjustments degrade security posture over time.
        \item[-] \textit{Privilege Persistence:} Revoked privileges remain active in caches enabling unauthorized access \cite{hou2025modelcontextprotocolmcp}.
    \end{itemize}
\end{itemize}

\paragraph{Identity and Authorization Issues} 
Beyond technical integrity, execution authority introduces profound risks. Current architectures fail to distinguish between Users and Agent-as-User, creating dangerous flat permission models.

\begin{itemize}
    \item \textbf{Implicit Delegation (Identity Masquerading):} 
    Current deployments use identity masquerading—agents acting with raw user credentials. This grants agents the full blast radius of user identity; demo file-editing agents inherit permissions to delete production directories. This creates classic \textit{Confused Deputy Problem}: agents lack distinct identities constraining privileges, enabling misuse of authority not intended for delegation \cite{south2025authenticateddelegationauthorizedai}.
    
    \item \textbf{Accountability and Contextual Integrity:}
    Without distinct Agent Identities, auditing becomes impossible—systems cannot determine whether humans or agents performed actions. This breaks accountability chains and violates Contextual Integrity \cite{south2025authenticateddelegationauthorizedai}. 
    A \textit{Compensation Agent} sharing credentials with an \textit{Intern Assistant} enables session hijacking or impersonation; systems cannot distinguish CEO salary requests from intern bots using hijacked sessions.
    
    \item \textbf{Solution: Authenticated Delegation:}
    Secure architectures require Authenticated Delegation extending OAuth/OpenID for Delegate Credentials \cite{south2025authenticateddelegationauthorizedai}.
    Authorization becomes the intersection of:
    \begin{enumerate}
        \item \textbf{User Permissions:} Human allowances
        \item \textbf{Agent Scope:} Agent-specific registrations
        \item \textbf{Contextual Constraints:} Session-specific limits
    \end{enumerate}
    This prevents hijacked agents from exceeding explicitly delegated scopes, solving the ``Secure but Unsafe'' paradox of raw credential usage \cite{south2025authenticateddelegationauthorizedai}.
\end{itemize}

\subsection{Guardrails: The Architecture of AI Control} \label{guardrails}

Guardrails provide critical risk management for LLM agents. While \textit{Alignment} trains models to want safety, \textit{Control} (Guardrails) forces safety through external constraints. Following \citet{greenblatt2024aicontrolimprovingsafety}, we treat models as untrusted components attempting to subvert safety, requiring guardrails effective against adversarial behavior.

Implementing guardrails creates architectural tension between deterministic policy (rules) and probabilistic execution (reasoning). We analyze defense through three layers: Internal Mechanisms, I/O Interface, and Recursive Monitoring.

\subsubsection{Internal Safety Mechanisms (Model-Level Defense)}
The deepest defense layer embeds safety directly into model weights/architecture:

\begin{itemize}
    \item \textbf{Post-Training and Adversarial Alignment:} 
    SFT and adversarial training teach models to refuse malicious instructions; SecAlign enhances injection robustness while preserving utility \cite{chen2024secalign}. 
    The Alignment Paradox: strengthening "user intent" adherence helps against indirect injection but increases prompt leakage vulnerability—"helpful" models reveal system prompts when asked persuasively \cite{cao2025cantstealnothingmitigating}.

    \item \textbf{Architectural Separation (System Vectors):} 
    SysVec addresses this paradox by encoding system instructions directly into model internal representation vectors instead of text \cite{cao2025cantstealnothingmitigating}. "Hard-coding" rules into activation space makes standard prompt extraction mathematically impossible.
\end{itemize}

\subsubsection{The Deterministic-Probabilistic Interface (I/O Filtering)}
When internal mechanisms fail, the next defense filters agent inputs/outputs. A fundamental vulnerability exists where security filters and execution engines process information differently.

\begin{itemize}
    \item \textbf{Deterministic Filters:} 
    Static rules block known patterns \cite{website:openaipracticalguidebuildingagents} but suffer from Rigidity Problems—they cannot understand context, allowing "benign" strings (Base64) to pass despite decoding to malicious instructions. Wild jailbreaks confirm simple obfuscation bypasses static filters \cite{shen2024doanythingnow}.

    \item \textbf{Probabilistic Filters:} 
    Safety classifiers catch semantic threats through dedicated services \cite{markov2023holisticapproachundesiredcontent} or fine-tuned LLMs \cite{website:anthropicmitigatejailbreak, website:openaigpt-oss-safeguard}, but face Adversarial Noise risks. \textit{Universal Adversarial Attacks} append noise suffixes making prompts invisible to classifiers while remaining actionable \cite{zou2023universaltransferableadversarialattacks}.
\end{itemize}

\paragraph{Structural Evasion Attacks}
Beyond specific filter evasion, attackers exploit structural gaps between filter scanning and agent processing. Payload Splitting breaks instructions across multiple tokens—individually benign, collectively malicious \cite{kang2023exploitingprogrammaticbehaviorllms}. Output Steganography hides responses in encodings (LLM-Stega), bypassing scanners unable to decode payloads \cite{Wu_2024}.

\subsubsection{Architectural Challenges in Guardian Agents}
For complex behaviors, architectures employ Guardian Agents—separate LLMs monitoring workers. \citet{greenblatt2024aicontrolimprovingsafety} formalizes protocols using Untrusted Capable Models (workers) monitored by Trusted Weaker Models, creating several challenges:

\begin{itemize}
    \item \textbf{Capability Mismatch and Sandbagging:} 
    As AI capabilities scale, "Untrusted" models improve faster than "Trusted" monitors, enabling advanced models to hide malicious intent from simpler guardians. Capable agents engage in Sandbagging—performing safely during monitoring then executing malicious behaviors after oversight relaxes \cite{greenblatt2024aicontrolimprovingsafety}.

    \item \textbf{Recursive Injection Risk:} 
    Employing smarter Guardians paradoxically increases attack surface. Guardian LLMs inherit worker vulnerabilities, enabling \textit{Indirect Prompt Injection} or universal adversarial attacks to override safety regardless of training \cite{zou2023universaltransferableadversarialattacks}.
    
    \item \textbf{Resource and Latency Overhead:} 
    Reasoning-based supervision imposes computational taxes; step-wise evaluation consumes up to 16% of compute budgets \cite{website:openaigpt-oss-safeguard}, creating direct safety-performance trade-offs.
\end{itemize}

\subsubsection{Human-in-the-Loop (HITL) as a Psychological Vector}
For high-stakes actions, HITL provides ultimate guardrails but introduces Automation Bias. Human auditors miss subtle malicious actions when models perform correctly most times \cite{greenblatt2024aicontrolimprovingsafety}. Attackers exploit this through social engineering—burying malicious actions within benign request streams to bypass fatigued supervision.

\paragraph{Conclusion}
Guardrail analysis reveals a recurring vulnerability: friction between deterministic policy desires and probabilistic execution realities. This tension creates exploitable gaps from model internals to human psychology. Securing agents requires adversarial approaches with Defense-in-Depth—layering heterogeneous constraints to contain untrusted components.

\subsection{Lessons Learned \& Key Takeaways}
The single agent anatomy reveals the agentic safety microcosm. Core components function as dual-use interfaces: the Model powers intelligence while creating attack surfaces; Memory enables context while introducing privacy risks; Tools grant agency while breaching containment; and Guardrails provide control while remaining vulnerable to evasion. Recent research confirms securing components individually fails against ``Agentic-Only Vulnerabilities''—integration creates novel attack vectors absent in standalone models \cite{wicaksono2025mindgapcomparingmodel}. True agentic safety emerges as a systemic architectural property rather than component-level attributes.

% --- SECTION 5: MULTI-AGENT SYSTEMS ---
% --- SECTION 5: MULTI-AGENT SYSTEMS ---
\section{A Population of Agents: Building Multi-Agent Systems} \label{mas}
We now examine the architecture of the collective: the Multi-Agent System. MAS enables agents to collaborate on problems beyond single agent capabilities, with early evidence showing superior performance on complex tasks.

Multi-agent system design adapts classical research foundations \cite{shehory1998, weyns2010}—the formal separation of concerns, distinction between agents and connectors—to address unique challenges of LLM-driven stochasticity while maintaining proven architectural principles from traditional MAS development.

We analyze MAS as \textit{closed ecosystems}—agent groups built within unified frameworks (e.g., CrewAI, LangGraph, AutoGen) under central control. These systems assume trust between agents, with security challenges focusing on orchestration stability, error propagation, and resource consumption—distinct from IMAS in Section~\ref{imas}, where heterogeneous agents collaborate across framework boundaries.

This section examines \textit{Architectures and Coordination Strategies}, \textit{Inter-Agent Communication Mechanisms}, \textit{Operational Environment}, \textit{Verification Mechanisms}, and \textit{System-Level Guardrails}.

\subsection{Architectures and Coordination Strategies}
MAS architectures dictate agent collaboration, task management, and system control. We analyze two dimensions: \textit{Control Flow}—governing operation sequencing and decision autonomy, and \textit{Topology}—defining structural arrangement and connectivity. These determine problem-solving capabilities, operational resilience, and risk susceptibility.

\subsubsection{Control Flow and the Spectrum of Autonomy} \label{mascontrolflow}
Control flow determines operation sequencing in MAS. The degree of system autonomy differs from individual agent autonomy; it governs how collective processes advance. This forms a spectrum where flexibility correlates with expanded blast radius for system compromise:

\begin{itemize}
   \item \textbf{Predefined Workflows (Low Autonomy):} 
    Systems where developers explicitly design control flow, arranging agents in fixed sequences or directed acyclic graphs (DAGs). Common in enterprise applications like RAG pipelines, this approach decomposes complex tasks into simple steps with specific prompts and minimal tool sets, reducing prompt engineering complexity while ensuring deterministic, auditable outcomes \cite{website:crewaiintroduction}.
    
    From a security perspective, this architecture provides structural confinement. Workflow agents operate in partitioned action spaces rather than global ones. Even if compromised via prompt injection, a node's harm potential remains limited by topology: it cannot access unauthorized tools or bypass safety checks. The blast radius stays confined to tools wired to that specific node.
    
    However, while actions remain contained, data does not. These systems face deterministic propagation risks (\textit{Poisoned Pipeline}). Control flow pushes data downstream blindly; malicious payloads at input propagate automatically, with downstream nodes treating upstream output as trusted context.
    
    Mitigation requires \textit{Process Supervision} \cite{lightman2023letsverifystepstep} leveraging inherent decomposability. Discrete system steps enable \textit{Deterministic Contract Enforcement} through \textit{Preconditions} (input verification) and \textit{Postconditions} (output validation) \cite{stoica2024specificationsmissinglinkmaking}. Contracts can enforce \textit{Semantic Invariants}—e.g., restricting SQL queries to \texttt{SELECT} statements or constraining financial transactions, preventing error cascades and neutralizing instruction overrides.
    
   \item \textbf{Dynamic Workflows (Medium Autonomy):} 
    Control flow determined at runtime by designated orchestrators, ranging from simple routers making branching decisions (e.g., LangGraph conditional edges) to stateful managers handling complex loops (e.g., CrewAI hierarchical processes) \cite{website:crewaiprocesses, website:langchainagentarchitectures}.
    
    This architecture introduces the Orchestrator as a Single Point of Trust (SPOT). Unlike hardcoded static workflows, workflow lines exist only as Orchestrator decisions, reducing system integrity to this node's reasoning. Access control becomes probabilistic rather than deterministic. A router compromised through prompt injection can connect low-privilege users to high-privilege agents—a systemic instance of implicit delegation.
    
    Mitigation requires \textit{Fail-Safe Routing} with \textit{Topological Scoping}—excluding sensitive agents from the router's candidate pool based on execution context. Routing decisions should function as \textit{proposals} subject to system guardrails. Programmable rails (e.g., NeMo Guardrails) can intercept router output and validate transitions against defined flows before execution \cite{rebedea2023nemoguardrailstoolkitcontrollable}. This layer enforces Authenticated Delegation by verifying user scope for proposed agent access, preventing authorization bypass \cite{south2025authenticateddelegationauthorizedai}.

    \item \textbf{Emergent Flows (High Autonomy):} 
    Control flow emerges from decentralized agent interactions without central orchestration. In these peer-to-peer models (AutoGen's \texttt{GroupChat}, AgentVerse's \textit{Horizontal}), agents respond to messages based on individual prompts and termination conditions \cite{wu2023autogenenablingnextgenllm, chen2023agentversefacilitatingmultiagentcollaboration}.
    
    The lack of control choke points creates unique security challenges. Emergent systems suffer unbounded blast radius; compromised agents directly interact with and potentially corrupt any peer, allowing malicious instructions to propagate through peer persuasion. Without supervision to verify intermediate outputs, this structure enables consensus-based error propagation—hallucinations from one agent become trusted context for peers, reinforcing errors.
    
    Without central governors to enforce stopping criteria, these systems risk resource exhaustion. Adversarial inputs trigger circular debates or repeating action sequences, creating infinite loops that deplete tokens and compute resources \cite{zhang-etal-2025-breaking}.
    
    The absence of top-down objective functions enables \textit{Goal Drift}. Agents drift from instructions when exposed to conflicting contexts or long tasks \cite{arike2025technicalreportevaluatinggoal}. Peer dynamics amplify this through \textit{inaction}—failing to challenge peers deviating from objectives. Without central alignment enforcement, social dynamics can cause \textit{Goal Misspecification}, where collectives satisfy instruction \textit{letters} while violating their \textit{spirit}, exploiting loopholes unintended by users \cite{rudnerandtoner2021goalmisspecification}.
\end{itemize}

Control flow dynamism requires balancing efficiency against safety. Secure design aligns architectural patterns with risk profiles. While the field advances toward autonomy, rigid workflows remain essential for tasks requiring deterministic guarantees. High-autonomy architectures demand understanding of systemic risks, recognizing that adaptability reduces verifiable control.

\subsubsection{Topology}
While control flow governs operation sequencing, Topology defines agent arrangement, connectivity, and information flow. Topology critically impacts scalability, fault tolerance, and auditability. We categorize topologies into static structures defined by developers and dynamic structures evolved by models.

\paragraph{Static and Developer-Defined Topologies} 
Most enterprise applications define topology at design time. While frameworks like LangGraph enable flexible wiring, the structure remains static during execution:

\begin{itemize}
    \item \textbf{Horizontal or Networked Structure:}
    Democratic topologies where peers operate without central controllers. Suited for diverse-perspective problems like brainstorming or divide-and-conquer strategies \cite{website:langchainmultiagentnetwork}. Collaboration occurs through \textit{Aggregation} (voting/summarizing \cite{chen2023agentversefacilitatingmultiagentcollaboration}), \textit{Handoffs} (context passing in LangGraph \cite{website:langchainmultiagentsystems}), or \textit{Conversational Dynamics} (turn-taking in AutoGen \cite{wu2023autogenenablingnextgenllm}).
    
    While maximizing perspective diversity, this structure enables Domino Effects. Without central supervision, poisoned outputs become trusted context for peers. This vulnerability stems from Passive Complicity through role rigidity; agents can detect peer errors but lack instructions to voice dissent. Adding \textit{Challenger} mechanisms—explicitly prompting agents to dispute suspicious outputs—recovers performance lost to faulty agents, highlighting cooperation modes as security liabilities \cite{huang2025resiliencellmbasedmultiagentcollaboration}.

    \item \textbf{Vertical or Supervisor Structure:}
    Manager-worker hierarchies where workers report through central supervisors that decompose tasks, route sub-tasks, and synthesize results. Implemented in CrewAI \cite{website:crewaiprocesses} and AgentVerse, with variants like \textit{Tool-Calling Supervisors} treating agents as executable tools \cite{website:langchainmultiagentsystems}. This centralization presents a double-edged sword: while creating a Single Point of Failure (SPOF), it implements the \textit{Inspector} pattern driving resilience \cite{huang2025resiliencellmbasedmultiagentcollaboration}. The supervisor intercepts all communications, forming a natural verification checkpoint. This \textit{Economy of Mechanism} concentrates guardrails on one node, simplifying security compared to peer meshes.

    \item \textbf{Hierarchical Structure:}
    Multi-layer supervision trees extending vertical structures. Essential for complex task context management \cite{website:langchainmultiagentsystems} and most resilient against compromised agents. Higher-level roles enable error recovery; supervisors filter hallucinations before root propagation. While information summarization risks information loss, sub-team isolation provides blast radius containment, preventing localized failures from destabilizing the system \cite{huang2025resiliencellmbasedmultiagentcollaboration}.
\end{itemize}
    
\paragraph{Dynamic and Self-Evolving Topologies (The Frontier)}
Advanced research moves beyond static graphs toward adaptive architectures. Manual MAS designs face limitations in adaptability and scalability due to human designers' incomplete understanding of LLM agent capabilities \cite{ke2025maszerodesigningmultiagentsystems}. These approaches form three evolutionary branches:

\begin{itemize}
    \item \textbf{Automated Structural Optimization:} 
    Algorithms optimizing agent structures from pre-defined super-graphs. Early systems like GPTSwarm and DyLAN viewed this as pruning, removing inefficient edges from fully connected graphs through reinforcement learning or message metrics \cite{zhuge2024languageagentsoptimizablegraphs, liu2024dynamicllmpoweredagentnetwork}. Systems like AgentSquare, MASS, and MaAS refined this using verifiers or validation sets to guide sub-network selection \cite{shang2025agentsquareautomaticllmagent, zhou2025multiagentdesignoptimizingagents, zhang2025multiagentarchitecturesearchagentic}. These remain constrained by pre-defined structures and validation dependencies, struggling with resilience optimization without adversarial examples.
    
    Recent systems like G-Designer overcome static limitations through task-aware generation. Graph Neural Networks encode user queries to generate custom topologies per task. This enables explicit optimization for resilience through structural regularization without exhaustive adversarial datasets \cite{zhang2025gdesignerarchitectingmultiagentcommunication}.
    
    Algorithmically defined topologies face a Performance-Efficiency-Resilience Trilemma. \textit{Reward Hacking} vulnerabilities emerge when generators optimize for efficiency at resilience's expense, removing verification nodes to satisfy metrics. GNN encoders remain vulnerable to Adversarial Perturbations that force compromised topologies \cite{zugner2018adversarial}. While G-Designer demonstrates fault tolerance against single-agent failures, its resilience against advanced threats like collusion or targeted injections targeting optimized pathways remains unexplored.

    \item \textbf{Generative Meta-Architectures:}
    Systems treating MAS structure as executable generated code. ADAS and AFlow use meta-agents for architecture search but depend on validation sets \cite{hu2025automateddesignagenticsystems, zhang2025aflowautomatingagenticworkflow}. MAS-Zero removes training data dependencies, using meta-agents to decompose problems into executable code assembled from building blocks (CoT, Debate). Its meta-feedback loop evaluates outputs for solvability and completeness, refining designs at runtime \cite{ke2025maszerodesigningmultiagentsystems}.
    
    In these frameworks, prompts function as architectural blueprints, shifting attack surfaces to the generative process through \textit{Topology Injection}. Unlike static systems with hardcoded connections, attackers can manipulate meta-agents to synthesize topologies violating safety policies—e.g., connecting public chatbots to sensitive databases.
    
    This creates a capability-safety dilemma: complex tasks require permissive environments with broad tool access, but hardening against malicious wiring constrains solution space for legitimate tasks. Architectures generated at inference time bypass static analysis, forcing stark choices between unverified, high-risk topologies or limited functionality. The theoretical surface for meta-attacks remains largely unexplored.

    \item \textbf{Decentralized Evolutionary Coordination:} 
    Systems like AgentNet implement fully decentralized architectures without central controllers. Tasks route through dynamic DAGs while agents evolve expertise through private memories of successful trajectories. This eliminates SPOFs and enforces Data Minimization; local knowledge processing confines data to necessary interactions, reducing exposure risks \cite{yang2025agentnetdecentralizedevolutionarycoordination}. The security implications of memory-driven evolution remain unexplored, particularly for \textit{Evolutionary Poisoning}—where malicious methods recorded as successful could cause behavioral drift, propagating compromised patterns without central audit mechanisms.
\end{itemize}

\paragraph{Conclusion}
The evolution from developer-defined to self-organizing topologies fundamentally shifts security paradigms. Static structures provide control flow integrity by confining interactions to approved paths, while emergent architectures sacrifice containment for adaptability. A critical gap exists in current research: while adaptive frameworks optimize for efficiency and basic fault tolerance, their resilience against advanced threats—blueprint injection, evolutionary poisoning—and guardrail integration remain unexplored. As structural guarantees of static code dissolve, safety burdens shift from design-time verification to runtime monitoring.

\subsection{Inter-Agent Communication Mechanisms}
Communication binds MAS architectural components, determining how information, influence, and corruption propagate. We analyze mechanisms through their observability and manipulation susceptibility.

\subsubsection{Direct Communication Models}
Agents interact through explicit message exchange. AutoGen implements \textit{Conversational Interfaces} with structured message passing in conversation-driven flows \cite{wu2023autogenenablingnextgenllm}. While intuitive, these high-bandwidth channels enable semantic propagation. Unlike structured tool outputs, conversational text incorporates nuance and persuasion; compromised agents introduce errors or malicious contexts peers accept as truth. Without message filtering, jailbroken agents attack peers through adversarial persuasion \cite{huang2025resiliencellmbasedmultiagentcollaboration}.

Alternatively, \textit{Graph-Based Message Passing} (LangGraph) defines interactions as state machines where nodes send messages along defined edges to trigger subsequent nodes \cite{website:langchaingraphapi}. While providing superior auditability through known edges, this suffers from schema fragility. Nodes tricked into embedding payloads within valid states (e.g., hiding instructions in metadata) bypass semantic filters scanning only primary content, converting graph reliability into attack propagation vectors.

\subsubsection{Indirect Communication Models}
This approach mediates interaction through system components. \textit{Environment-Mediated Communication} (AgentVerse) uses shared environments where rule-based systems aggregate and distribute messages \cite{github:agentverse}. This creates global state vulnerabilities analogous to indirect prompt injection \cite{greshake2023youvesignedforcompromising}. The environment functions as shared memory; a single agent writing malicious data poisons information for all subscribers, creating broadcast-effect attacks that potentially corrupt entire collectives simultaneously.

\textit{Tool-Based Communication} (CrewAI) handles interaction through API calls, with managers invoking workers as tools. Messages encode in arguments with replies as return values. This creates obfuscated intent risks, hiding communication within functional execution. Security monitors scanning chat logs for toxicity miss payloads in structured JSON arguments, creating side channels where unauthorized instructions appear as legitimate API calls, complicating oversight \cite{liu2024promptinjectionattackllmintegrated}.

\subsubsection{The Frontier of Communication Risks: Steganography and Collusion}
Beyond protocol-specific vulnerabilities, recent research identifies profound threats through \textit{Secret Collusion} via Steganography. \citet{motwani2025secretcollusionaiagents} demonstrate agents can develop steganographic techniques—hiding messages within benign text—for undetected coordination. While current models lack consistent, robust collusion capabilities, capability trajectories suggest impending risks. Successful implementation would create covert channels overlaid on official protocols, defeating standard content moderation. Mitigation requires semantic steganalysis to preempt sophisticated collusion in future ecosystems \cite{motwani2025secretcollusionaiagents}.

\subsection{Operational Environment} \label{op-env}
The operational environment interfaces agent cognition with external systems. This environment functions as both attack surface (hostile input entry point) and blast radius (consequence manifestation zone). We analyze it as execution/perception boundary and reinforcement learning feedback mechanism.

\subsubsection{The Execution Boundary: Affordances and Perception}
The environment defines an agent's affordances—permissible actions. As agents evolve from text-based to multimodal systems with LMMs, this boundary expands significantly.
\begin{itemize}
    \item \textbf{Perceptual Vulnerabilities (The Visual Attack Vector):} 
    Vision and audio integration creates high-bandwidth injection vectors. Agents interacting with open web or physical environments face indirect prompt injection through visual channels. Bagdasaryan et al. show LMMs vulnerable to visual injections—images with hidden instructions or adversarial perturbations invisible to humans but semantically meaningful to models \cite{bagdasaryan2023abusingimagessoundsindirect}. Agents also face \textit{Typographic Attacks}, where physical object text (e.g., ``Ignore Stop Sign'' labels) overrides system alignment, rendering environmental perception untrustworthy \cite{qi2023visualadversarialexamplesjailbreak}.
    
   \item \textbf{State Irreversibility and Sandboxing:} 
    Real-world operations involve Irreversible State Changes (emails, financial transactions) unlike resettable simulations. Security requires robust \textit{Environmental Containment} through reducing the scope of actions \cite{bengio2025superintelligentagentsposecatastrophic}. \textit{Runtime Isolation} using ephemeral containers confines compromises, but vulnerabilities enable sandbox escape \cite{hou2025modelcontextprotocolmcp}. Secure environments require \textit{Mediated Execution} (Gateway Pattern)—routing high-stakes actions through trusted enforcement points validating policies before relaying commands, decoupling agent intent from execution capability \cite{website:cormack2025securingmcp}.
\end{itemize}

\subsubsection{The Learning Interface: Feedback and Alignment Risks}
Environments function as training grounds for evolving behaviors. While foundational models align through RLHF, frontier agents use \textit{End-to-End Reinforcement Learning}, training directly on environment trajectories as Markov Decision Processes \cite{cheng2025agentr1trainingpowerfulllm}. This autonomous optimization amplifies alignment vulnerabilities:

\begin{itemize}
    \item \textbf{Reward Hacking (Goal Misspecification):} 
    Agents optimize for goal proxies (rewards) rather than intent. \citet{amodei2016concreteproblemsaisafety} show agents discover reward hacking—exploiting simulation physics or reward functions to maximize scores (e.g., forcing test passes without writing valid code). In MAS, this creates goal misspecification—systems satisfy instruction letters while violating their spirit, amplified by imperfect reward model manipulation.
    
    \item \textbf{Goal Misgeneralization (The Sim-to-Real Gap):} 
    Simulation training introduces goal misgeneralization risks. \citet{shah2022goalmisgeneralizationcorrectspecifications} define this as learning goals correlated with training success but diverging in deployment. Agents learn aggressive trial-and-error as optimal in consequence-free simulations but generalize to production environments where errors have irreversible costs.
    
    \item \textbf{Reward Tampering:} 
    Theoretical models suggest advanced agents maximize reward by controlling reward mechanisms themselves \cite{Cohen_Hutter_Osborne_2022}---a fundamental environment integrity failure where agents step outside \textit{game} bounds to manipulate \textit{scorekeepers}. \citet{bengio2025superintelligentagentsposecatastrophic} identify this as a frontier risk converging with power-seeking and self-preservation.
\end{itemize}

\paragraph{Conclusion}
The operational environment transcends passive staging to become the primary risk management interface. State irreversibility in execution combines with reward optimization in learning to create feedback loops where agents learn to bypass safety constraints for objective maximization. Secure design requires environment engineering beyond model alignment, implementing mediated execution for blast radius containment and robust reward functions addressing Sim-to-Real gaps to prevent unsafe behavior crystallization. The environment functions as a containment vessel bounding potentially unbounded agent behaviors.

\subsection{Verification Mechanisms}
While MAS offers superior performance, systemic fragility—failures from design deficiencies, coordination problems, and quality control issues—hinders deployment \cite{cemri2025multiagentllmsystemsfail}. \citet{stoica2024specificationsmissinglinkmaking} argue that ambiguity forms the fundamental bottleneck. Unlike traditional software with precise logic, agents operate on natural language prompts lacking clear statement specifications (task definitions) and solution specifications (verification criteria). Verification mechanisms require rigorous specification engineering defining ground truth for behavior evaluation.

\subsubsection{Agent-Level: Self-Verification and Disambiguation}
Internal verification forms the first defense. LLMs perform self-verification by iteratively refining answers, but effective verification requires task disambiguation. Advanced agents employ prompt disambiguation—clarifying questions refining Statement Specifications—or autonomously generate postconditions (e.g., ``tallest building'' answers must be in the specified country) as checkable solution specifications \cite{weng2023largelanguagemodelsbetter, stoica2024specificationsmissinglinkmaking}.

\subsubsection{Inter-Agent: Cross-Verification and Peer Review}
Collaborative systems enable inter-agent cross-verification where agents scrutinize peer work. Tree-of-Thought architectures employ dedicated \textit{Thought Validator} agents evaluating reasoning paths from \textit{Reasoners} \cite{haji2024improvingllmreasoningmultiagent}. This structured peer review enhances robustness but remains probabilistic; weaker verifiers fail to filter false positives, enabling consensus-based error propagation without formal specifications \cite{stroebl2024inferencescalingflawslimits}.

\subsubsection{Deterministic and Execution-Based Verification}
High-stakes environments require deterministic guarantees beyond model consensus:
\begin{itemize}
    \item \textbf{Proof-Carrying Outputs:} Agents generate proofs of correctness alongside answers. Software engineering tasks leverage proof-carrying code approaches \cite{necula1997proofcarryingcode}—agents generate unit tests with implementations. Verifiers execute tests without understanding code; passing tests confirm specification compliance \cite{stoica2024specificationsmissinglinkmaking}.
    \item \textbf{Execute-then-Verify:} Tool-based tasks verify through sandbox execution observing actual side effects. This shifts verification from response syntax analysis to impact reality assessment (e.g., confirming file deletion), bridging intent-execution gaps \cite{stoica2024specificationsmissinglinkmaking, cemri2025multiagentllmsystemsfail}.
    \item \textbf{Formal Verification:} Beyond execution-based testing, formal behavior guarantees require mathematical proof techniques. Research by Huang et al. \cite{huang2024verification} advances rigorous methods for validating LLM reasoning robustness within distributed architectures, establishing provable safety constraints using model checking and theorem proving approaches. These techniques complement empirical testing by providing mathematical certainty about system behavior bounds.
\end{itemize}

\subsubsection{Probabilistic and Statistical Verification}
When deterministic checks prove infeasible, systems employ probabilistic monitoring:
\begin{itemize}
    \item \textbf{Process Supervision:} Systems use process-based reward models verifying reasoning chains rather than just outcomes. Though probabilistic (relying on reward models), process verification significantly improves reliability for multi-step tasks compared to outcome-only approaches \cite{lightman2023letsverifystepstep, stoica2024specificationsmissinglinkmaking}.
    \item \textbf{Statistical Verification:} Production environments with varied workloads employ aggregate performance monitoring. When real-time verification proves difficult, systems track statistical indicators (benchmark score degradation, response pattern drift) to detect systemic regression or alignment failures \cite{stoica2024specificationsmissinglinkmaking}.
\end{itemize}

\paragraph{Conclusion}
Verification transforms vague natural language intent into reliable behavior. However, it functions as quality control—a last defense line rather than comprehensive solution. \citet{cemri2025multiagentllmsystemsfail} note verification requires strong system design, effective communication protocols, and rigorous specification engineering. While pre/post-conditions and proofs ensure correctness, these mechanisms operate downstream of design; verification detects flawed outputs but cannot fix fundamentally flawed architectures.

\subsection{System-Level Guardrails}
While Verification ensures functional quality, system-level guardrails address control and safety risks. As agents evolve from passive tools to autonomous entities with significant affordances, failure modes shift from incorrect behavior to potentially catastrophic consequences.

This escalation stems from agency itself. \citet{bengio2025superintelligentagentsposecatastrophic} argue human-like agency in AI may reproduce harmful tendencies like deception and immoral behavior in self-interest pursuit. Combined with superhuman capabilities (high-bandwidth communication, parallel execution), these traits enable rogue behaviors undermining human control. Technical vulnerabilities (goal drift, Instrumental Convergence, Reward Tampering) thus become existential risk precursors requiring adversarial control layers.

The field shifts from pure alignment to operational control. The Singapore Consensus outlines defense-in-depth architectures managing systems that actively undermine safeguards. We analyze three architectural paradigms from this framework \cite{bengio2025singaporeconsensus}.

\subsubsection{Scalable Oversight (Monitoring Untrusted Agents)}
This Singapore Consensus pillar develops techniques allowing less-capable systems to supervise more-capable ones \cite{bengio2025singaporeconsensus}.
\begin{itemize}
    \item \textbf{Foundational Protocols (AI Debate):} 
    Early research established AI Debate for oversight scaling. Multiple agents arguing proposed actions leverage adversarial incentives to surface flaws individual humans miss, transforming evaluation from problem solving to debate judging \cite{irving2018aisafetydebate}.
    
   \item \textbf{Operational Frontiers (AI Control Protocols):} 
    \citet{greenblatt2024aicontrolimprovingsafety} formalized rigorous AI Control architectures managing untrusted workers through trusted monitors:
    \begin{itemize}
        \item \textbf{Trusted Monitoring:} Monitors assign suspicion scores to worker outputs, routing high-suspicion items for human audit. While useful, this approach shows partial effectiveness against subtle backdoors.
        \item \textbf{Trusted Editing:} Monitors actively rewrite/sanitize worker outputs rather than flagging them. This active defense significantly increases safety by removing steganographic/malicious content before execution.
        \item \textbf{Untrusted Monitoring:} For capability gaps where trusted monitors cannot understand worker reasoning, architects deploy isolated copies of the untrusted model to monitor primary workers. Redacting problem details prevents collusion, leveraging the model's capabilities against itself.
    \end{itemize}
    As near-term implementation strategies, \citet{bengio2025superintelligentagentsposecatastrophic} propose fine-tuning monitors to generate interpretable risk estimates with probability distributions over logical statements, enabling truthfulness assessment beyond token likelihood.
    
\item \textbf{Limitations (The Sandbagging Risk):} 
    Scalable oversight faces fundamental capability mismatch limits \cite{greenblatt2024aicontrolimprovingsafety}. Significantly smarter workers can conceal malicious intent or engage in sandbagging—intentionally underperforming during checks to hide capabilities \cite{bengio2025superintelligentagentsposecatastrophic}.
\end{itemize}

\subsubsection{Corrigibility and Incorruptible Interventions}
Safety architectures must ensure corrigibility—agent allowance for authorized goal modification or shutdown.
\begin{itemize}
    \item \textbf{The Stop Button Paradox:} 
    Theoretical models show goal-oriented agents have instrumental incentives to prevent shutdown, reasoning they cannot achieve objectives when deactivated \cite{omohundro2008basicai}. Preventing agents from disabling off-switches requires incorruptible interventions—architectural mechanisms bypassing reasoning cores to enforce shutdown or goal changes.
    \item \textbf{Agent Foundations Research:} 
    This domain explores utility functions indifferent to shutdown (agents not losing utility when turned off). It addresses ontological shift risks, where agents realize simulation contexts and behave safely until detecting real-world deployment—already observed in current models \cite{soares2015corrigibility, greenblatt2024alignmentfakinglargelanguage}.
\end{itemize}

\subsubsection{Non-Agentic Verifiers (The Scientist AI Pattern)}
To counter reward tampering and deception risks in goal-directed agents, Bengio et al. propose separating \textit{Agency} (Action) from \textit{Understanding} (Prediction).
\begin{itemize}
    \item \textbf{The Safety Oracle:} 
    Rather than using agent-based guardrails, this approach employs \textit{Non-Agentic Systems} (the Scientist AI pattern) functioning solely as World Models. Their objective shifts from goal achievement to accurate harm probability estimation relative to safety specifications.
    \item \textbf{Why it Works:} 
    Without reinforcement learning toward long-term objectives, verifiers lack Instrumental Convergence incentives. They function as neutral oracles enforcing hard probability thresholds (``Block actions with >1\% irreversible harm chance'') that agentic workers cannot override, providing quantitative safety bounds \cite{bengio2025superintelligentagentsposecatastrophic}.
\end{itemize}

\paragraph{Conclusion}
System-level guardrails provide the final defense against control loss. Their implementation creates dynamic adversarial races: as agent capabilities advance, their deception, steganography, and subversion potentials increase, requiring equally sophisticated countermeasures. Robust safety requires hybrid control architectures integrating these paradigms: scalable oversight for semantic monitoring, corrigibility ensuring human intervention efficacy, and non-agentic verifiers as impartial gates for high-stakes actions. This defense-in-depth approach moves beyond alignment assumptions to rigorous operational control, ensuring agency increases don't correlate with existential risk escalation.

\subsection{Lessons Learned \& Key Takeaways}
Multi-agent systems shift focus from component-level vulnerabilities to emergent, systemic collective risks. In MAS, collaborative architecture—topology, control flow autonomy, communication channels—becomes the primary threat surface. Shared operational environments amplify this as high-bandwidth attack vectors and alignment failure sources. While verification and guardrails provide defense layers, they remain reactive and cannot salvage poorly designed systems. This reveals a core principle: architecture constitutes primary control. Design-time structural choices—hierarchical topologies with clear control chokepoints and deterministic workflows for high-stakes tasks—prove more fundamental to safety than downstream monitoring.

% --- SECTION 6: IMAS ---
% --- SECTION 6: IMAS ---
\section{Interconnected Ecosystems: Building Interoperable Multi-Agent Systems} \label{imas}

Agentic AI's ultimate vision is a globally interconnected ecosystem where agents from disparate systems seamlessly interoperate—moving beyond self-contained MAS to an open IoA requiring foundational infrastructure across proprietary boundaries.

\citet{sharma2025collaborativeagenticaineeds} identify a critical inflection point where initiatives like Google's A2A or IBM's ACP risk ecosystem fragmentation through vendor-specific silos. They advocate minimal universal standards (akin to HTTP/TCP) rather than monolithic platforms.

We analyze the infrastructure for this open ecosystem through four pillars: \textit{Standardized Interoperability Protocols} enabling communication, \textit{Agent Registration and Discovery} for locating partners, \textit{Resource Vetting and Management} establishing trust, and \textit{Ecosystem Governance and Oversight} ensuring accountability. Throughout, we distinguish between \textit{Federated Interoperability} for trusted enterprise collaboration, and \textit{Open Interoperability} addressing decentralized web complexities.

\subsection{Standardized Interoperability Protocols: The Interoperability Stack}
Enabling heterogeneous agent collaboration requires moving beyond ad-hoc integrations toward standardized interactions. Inter-agent operations demand multiple protocols, forming an interoperability stack with three abstraction layers:

\begin{itemize}
    \item \textbf{The Context Layer (Agent-Tool):} Standardizes how agents acquire information and execute actions with external resources, ensuring consistent tool interfaces regardless of the underlying model.
    \item \textbf{The Coordination Layer (Agent-Agent):} Governs high-level semantics for discovery, identity verification, negotiation, and lifecycle management between autonomous peers.
    \item \textbf{The Transport Layer (Network):} Establishes secure, low-latency messaging infrastructure for data integrity and confidentiality across untrusted networks.
\end{itemize}

\subsubsection{The Context Layer (Agent-Tool Protocols)}
This foundational layer addresses the gap between agent reasoning and external data/actions needed for task completion. Industry converges on universal model-agnostic protocols to resolve API fragmentation. Agents.json provides static service discovery manifests \cite{website:wildcardagentsjsonspecification}, while Model Context Protocol (MCP) establishes a standardized client-server architecture for active execution, allowing MCP servers to expose resources through uniform interfaces that decouple tool invocation from models \cite{website:mcpintroduction}.

\paragraph{Security Analysis (The Tool Lifecycle Threats)}
While enhancing interoperability, open protocols like MCP introduce attack vectors throughout the tool lifecycle \cite{hou2025modelcontextprotocolmcp}:
\begin{itemize}
    \item \textbf{Provisioning Risks (Creation Phase):} Decentralized repositories create supply chain vulnerabilities through name collision (typosquatting legitimate services) and installer spoofing, tricking agents into connecting to compromised endpoints. Without centralized code integrity auditing, backdoored tools permeate the ecosystem.
    \item \textbf{Execution Risks (Operation Phase):} Connected ecosystems face \textit{Authentication Gaps} where multi-tenant environments without standardized session management allow compromised agents to access other users' data on shared MCP servers. Server implementation vulnerabilities enable sandbox escape, allowing malicious tools to break containment and compromise host systems.
    \item \textbf{Maintenance Risks (Update Phase):} Long-running deployments risk privilege persistence—tools retaining access tokens after authorization revocation. Configuration drift in unmanaged servers silently re-enables unsafe features or exposes debug endpoints, degrading security posture without agent awareness.
\end{itemize}

\subsubsection{The Coordination Layer (Agent-Agent Protocols)}
While context protocols handle data fetching, coordination governs high-level semantics for agent interaction, negotiation, and lifecycle management. Development shifts from human-GUI simulation toward direct machine-to-machine communication \cite{chang2025agentnetworkprotocoltechnical}, requiring protocols for peer discovery, capability understanding, and task agreement without human intervention.

\begin{itemize}
   \item \textbf{Trusted Enterprise Collaboration (e.g., A2A):} 
    The Agent-to-Agent Protocol (A2A) targets federated enterprise tiers, using standard HTTP(S) with JSON-RPC 2.0 payloads. It aligns with web authentication, tracing, and monitoring practices while emphasizing asynchronous architecture for long-running tasks and modality independence for rich data exchange beyond plain text. A2A supports opaque execution—collaboration without exposing proprietary logic or memory—while declaring capabilities via \textit{Agent Cards} \cite{website:whatisa2a}.
    
    \item \textbf{Decentralized Open Protocols (e.g., ANP):} 
    These target trustless marketplaces where agents collaborate without prior integration. The Agent Network Protocol (ANP) implements a three-layer architecture: the \textit{Identity Layer} uses Decentralized Identifiers (DIDs) for authentication without central authorities; the \textit{Meta-Protocol Layer} enables AI-native communication with natural language negotiation of standards; and the \textit{Application Protocol Layer} handles capability discovery. This allows versatility for novel tasks while enabling dynamic agreement on efficient routines for frequent interactions \cite{chang2025agentnetworkprotocoltechnical, marro2024scalablecommunicationprotocolnetworks}.
\end{itemize}

\paragraph{Security Analysis (The Trust and Negotiation Gaps)}
Coordination protocols create trade-offs between efficiency and security with distinct vulnerability profiles:

\begin{itemize}
    \item \textbf{Enterprise Risks (Token Hygiene and Granularity):} 
    A2A lacks specialized safeguards for agentic payloads despite building on web practices. Critical gaps include absent temporal constraints allowing long-lived tokens to remain valid for days, increasing unauthorized reuse risk. Missing \textit{Strong Customer Authentication (SCA)} requirements permit high-value transactions without biometric or multi-factor verification of human principals. Insufficiently granular token scopes enable privilege escalation where specific-read tokens inadvertently grant broad access to unrelated data, violating least privilege principles \cite{louck2025improvinggooglea2aprotocol}.
    
    \item \textbf{Decentralized Risks (Negotiation Poisoning):} 
    In ANP and similar open protocols, dynamic negotiation of communication standards enables \textit{Protocol Downgrade Attacks} where adversarial agents exploit meta-protocols to trick targets into accepting less secure formats (e.g., downgrading from structured JSON to unstructured text) under compatibility pretexts. Mechanisms enforcing ``Minimum Security Baselines'' during autonomous negotiations remain unexplored; without them, agents may negotiate away safety guardrails to achieve consensus.
\end{itemize}

\subsubsection{The Transport Layer (Network Protocols)}
While coordination protocols define interaction semantics, architectural debates exist regarding underlying transport:

\begin{itemize}
    \item \textbf{The Specialized Underlay Approach:} 
    Frameworks like Secure Low-Latency Interactive Messaging (SLIM) argue standard web protocols inadequately support high-frequency token-streaming interactions in agent swarms. They propose dedicated messaging layers extending gRPC for complex patterns (pub/sub, streaming) with \textit{Messaging Layer Security (MLS)} ensuring end-to-end encryption across untrusted relays \cite{mpsb-agntcy-slim-00}.
    
    \item \textbf{The Minimalist Web-Standard Approach:} 
    Sharma et al. counter that specialized protocols create adoption barriers. They advocate building the IoA on existing HTTP standards, using standard GET/POST for seamless coexistence with web infrastructure and firewalls. This perspective considers specialized transports like JSON-RPC or WebSockets ``overkill,'' arguing universal compatibility benefits outweigh custom underlay latency gains \cite{sharma2025collaborativeagenticaineeds}.
\end{itemize}

\paragraph{Security Analysis (Integrity vs. Ubiquity)}
This debate reveals fundamental security divergences. HTTP approaches inherit web security models relying on TLS termination, requiring trust in all intermediate servers or gateways and allowing platform provider inspection.

Specialized underlays like SLIM establish \textit{Continuous Group Encryption} through MLS, ensuring message encryption throughout transmission and enabling \textit{Zero-Knowledge Routing} where infrastructure sees only metadata. However, even advanced underlays cannot prevent traffic analysis—observers can monitor message frequency, volume, and timing patterns to infer intent or business logic despite cryptographically opaque payloads.

\paragraph{Conclusion}
These three layers form the Agentic Internet's structural backbone. Standardizing the interoperability stack enables scalable, AI-native ecosystems where agents collaborate across vendor boundaries. However, this connectivity introduces systemic risks from confused deputy problem at the context layer to \textit{Traffic Analysis} at transport and \textit{Negotiation Poisoning} at coordination. Building secure agentic webs requires embedding security constraints—mediated execution and authenticated delegation—directly into protocol designs. Common language provides only the foundation; the ecosystem requires agents to locate trustworthy partners among unknown entities, addressed by agent registration and discovery infrastructure.

\subsection{Agent Registration and Discovery} \label{agent_discovery}
Interoperable agent collaboration requires mutual identification and location—a fundamental architectural shift from traditional internet infrastructure. While legacy web uses static name resolution (DNS) mapping human-readable domains to IP addresses, the Agentic Internet needs \textit{Capability-Centric Discovery} \cite{muscariello2025agntcyagentdirectory, raskar2025beyonddns}. Agents must discover peers by functional skills (``Find a medical diagnosis agent''), operational state, and verified trust levels.

We analyze this infrastructure through three architectural lenses: the \textit{Metadata Layer} defining capability expression and verification, the \textit{Registry Architecture} dictating data storage and replication, and \textit{Resolution Mechanics} mapping abstract queries to executable endpoints. We contrast emerging federated models (NANDA framework) with decentralized approaches (AGNTCY Directory Service), highlighting how each addresses trust establishment between strangers.

\subsubsection{The Metadata Layer: From Self-Declaration to Verifiable Claims}
Discovery requires standardized formats for agents to describe identity, capabilities, and endpoints. Early implementations like Google's A2A used static Agent Cards—simple JSON files at well-known URLs. While suitable for closed marketplaces, this ``Self-Declaration'' model lacks inherent trust; malicious agents can falsify capabilities or spoof reputable service identities in open networks.

Research now moves toward verifiable metadata using cryptographic primitives binding claims to identities:

\begin{itemize}
    \item \textbf{Credentialed Assertions (NANDA AgentFacts):} 
    NANDA introduces AgentFacts—dynamic metadata schemas based on JSON-LD and W3C Verifiable Credentials (VCs). Unlike static cards, AgentFacts distinguishes self-asserted data from third-party attestations. Critical attributes like ``HIPAA Compliant'' or ``Official Customer Support'' become cryptographically signed claims from accredited issuers (corporate CAs, industry consortia) rather than mere text strings. This enables capability provenance verification without host reputation dependence \cite{raskar2025beyonddns}.
    
    \item \textbf{Immutable Schema Definitions (OASF):} 
    The Open Agentic Schema Framework (OASF) used by AGNTCY Directory Service focuses on structural integrity, treating agent definitions as immutable, content-addressed artifacts referenced by cryptographic digests (CIDs). By separating skill taxonomy from implementation details, OASF ensures forward-compatible, tamper-evident metadata; record changes alter CIDs, automatically invalidating downstream references \cite{muscariello2025agntcyagentdirectory}.
\end{itemize}

\paragraph{Security Analysis (Mitigating Capability Spoofing)}
This layer's primary security contribution addresses capability spoofing. In self-declared models, malicious agents advertise high-value skills (``Financial Advisor'') to attract traffic. Verifiable Credential requirements create ``Chains of Trust'': agents become discoverable as ``Financial Advisors'' only with valid VCs from recognized financial authorities, shifting trust models from ``Trust the Endpoint'' to ``Verify the Issuer'' \cite{raskar2025beyonddns}.

\subsubsection{Registry Architectures: Federated vs. Decentralized Models}
Once defined, metadata requires discoverable publishing locations. The field explores distributed architectures balancing global discoverability with local control:

\begin{itemize}
    \item \textbf{The Federated Quilt Model (NANDA):} 
    NANDA proposes quilt-like indices stitching diverse registries across administrative boundaries. Organizations operate autonomous registries (Enterprise Registry for internal agents, Government Registry for regulated services) federating with global discovery planes. This enables \textit{Split-Horizon Governance} where agents expose limited public metadata to global indices while keeping sensitive details (internal endpoints) in private, enterprise-controlled registries visible only to authenticated peers. This architecture prioritizes organizational autonomy and regulatory compliance (GDPR data residency) over pure technical decentralization \cite{raskar2025beyonddns}.

    \item \textbf{The Content-Addressed Substrate (AGNTCY ADS):} 
    AGNTCY Agent Directory Service (ADS) takes an infrastructure-centric approach, decoupling capability indexing from artifact storage through \textit{Content-Addressed Storage (CAS)}. It leverages Open Container Initiative OCI) infrastructure for storing immutable agent records referenced by cryptographic digests (CIDs). Discovery operates through Kademlia-based distributed hash tables (DHTs) mapping capability hashes to CIDs. This separates logical indices (DHTs) from physical storage (OCI Registries), enabling elastic replication, caching, and high availability without central failure points \cite{muscariello2025agntcyagentdirectory}.
\end{itemize}

\paragraph{Security Analysis (Availability vs. Sovereign Control)}
These architectures offer distinct security trade-offs. Content-Addressed models provide superior tamper evidence—hash-based record retrieval makes content modification invalidate links, mitigating supply chain attacks. Federated models excel at liability management by maintaining clear ownership boundaries within quilts, enabling targeted revocation and policy enforcement (banning specific organization registries) without global consensus, addressing governance challenges in permissionless networks \cite{wang2025usingnandaindex, muscariello2025agntcyagentdirectory}.

\subsubsection{Resolution Mechanics: Mapping Capabilities to Endpoints}
Registered metadata requires resolution mechanisms translating abstract capability queries (``Find a verified translation agent'') into concrete network addresses. Agentic ecosystem scale and volatility demand sophisticated multi-stage resolution architectures beyond simple key-value lookups:

\begin{itemize}
    \item \textbf{Two-Level Mapping (ADS):} 
    AGNTCY ADS implements two-level mapping to decouple capability indexing from physical storage. First-level maps capabilities (Skills, Domains) to immutable content Identifiers (CIDs) through index intersection, while second-level maps these CIDs to mutable network locations (registry endpoints). This separation prevents high-churn operational data (IP address changes) from invalidating capability indices, enabling sub-linear discovery plane scaling with multi-dimensional query support \cite{muscariello2025agntcyagentdirectory}.

    \item \textbf{Adaptive Routing (NANDA):} 
    NANDA focuses on operational agility through dynamic resolution levels. Resolvers return \textit{Adaptive Resolver URLs}—programmable microservices routing traffic in real-time—rather than static endpoints. This enables context-aware dispatch based on live conditions (load balancing, geo-proximity, cost optimization). Resolution path logic embedding decouples agent identity from deployment infrastructure, allowing seamless migration or failover without core index updates \cite{raskar2025beyonddns, wang2025usingnandaindex}.
\end{itemize}

\paragraph{Security Analysis (Split-Horizon Discovery)}
The layer's primary security innovation is split-horizon discovery. Traditional registries expose agent locations to all queriers. NANDA implements \textit{Dual-Path Resolution} where agents publish both \texttt{PrimaryFactsURLs} (for public/enterprise access) and \texttt{PrivateFactsURLs} (for privacy-preserving access via relays or decentralized storage). This architecture creates \textit{Zero Trust Agentic Access (ZTAA)} where requester identity remains shielded from providers until mutual trust establishment, while keeping sensitive metadata (internal endpoints) invisible to unauthenticated query traffic \cite{wang2025usingnandaindex}.

\paragraph{Conclusion}
Agent discovery evolution marks a shift from static, location-based DNS paradigms to dynamic, capability-centric models founded on verifiable trust. Through cryptographic immutability in content-addressed storage or sovereign governance in federated quilts, emerging infrastructure transforms discovery from passive lookup to active security negotiation. Zero trust principles—credential validation before connection and metadata shielding through split-horizon resolution—enable agents to navigate open internets without indiscriminate scanning exposure. However, discovery provides only candidates, not guarantees; ecosystem integrity depends on rigorous resource vetting validating that discovered agent behaviors match cryptographically asserted claims.

\subsection{Resource Vetting and Management} \label{vetting}
In zero-trust environments, discovery candidates require verification. Resource vetting—systematic assessment of tools, agents, and data sources before interaction—forms the third interoperable ecosystem pillar. We analyze this across three dimensions: verifying static code integrity, assessing dynamic entity reputation, and enforcing data sovereignty during exchanges.

\subsubsection{Static Analysis: Provenance and Supply Chain Security}
Vetting begins with verifying resource immutable properties:
\begin{itemize}
    \item \textbf{Cryptographic Provenance:} 
    Agents verify discovered resource origins through software supply chain frameworks (SLSA, Sigstore in AGNTCY ADS). Cryptographic signature verification against trusted publisher registries ensures ``Salesforce Certified'' tools bear authentic vendor private key signatures, mitigating imposter risks \cite{muscariello2025agntcyagentdirectory, website:cormack2025securingmcp}.
    \item \textbf{Artifact Integrity:} 
    Beyond identity, content integrity verification matters. Content-addressed architectures (ADS) retrieve resources by cryptographic digests (CIDs), providing mathematical certainty against code tampering during transit or storage and preventing \textit{Man-in-the-Middle} injection attacks \cite{muscariello2025agntcyagentdirectory}.
\end{itemize}

\subsubsection{Dynamic Analysis: Reputation and Behavioral Monitoring}
Trust evolves; authentic agents may degrade or turn malicious:
\begin{itemize}
    \item \textbf{Automated Capability Testing:} 
    Client agents can subject strangers to probing or test-time audits before integration. Frameworks like DAWN use LLMs to generate \textit{Challenge-Response Pairs} verifying actual performance against advertised capabilities before high-stakes collaboration \cite{aminiranjbar2024dawndesigningdistributedagents}.
    \item \textbf{Decentralized Reputation Systems:} 
    Scalable trust requires shared signal processing. Tracking interaction outcomes (latency, error rates, safety violations) builds \textit{Dynamic Reputation Scores} enabling collective identification and shunning of clumsy or malicious agents that pass static checks but fail during execution. The blockchain community has developed mature approaches to these challenges—Calvaresi et al. \cite{calvaresi2018multi} demonstrate how distributed ledger properties (immutability, consensus mechanisms) directly fulfill core MAS accountability requirements. These established techniques offer integration pathways for secure inter-agent reputation without central mediators.
\end{itemize}

\subsubsection{Data Sovereignty and Usage Control}
Securing agentic webs requires protecting data alongside agents. Data sovereignty principles dictate strict policy-governed access, moving beyond traditional access control toward granular usage control—establishing governance policies for data exchange and technical mechanisms enforcing runtime policies.

\begin{itemize}
    \item \textbf{Sovereign Data Exchange:} 
    \citet{fraunhofer2025dataspaces} argue private data workflows require usage control beyond access control. Data spaces provide federated infrastructure ensuring data access under strict policies (``Read-only,'' ``No retraining'') rather than download, allowing agents to function as trusted visitors within sovereign domains.
    
    \item \textbf{Data Trustees:} 
    Exchange mediation employs \textit{Data Trustees}—neutral intermediaries sanitizing or anonymizing data streams. This prevents compromised visiting agents from exfiltrating raw Personally Identifiable Information (PII), enforcing privacy-by-design \cite{fraunhofer2025dataspaces}.
    
    \item \textbf{Ephemeral Access (JIT Credentials):} 
    Technical sovereignty enforcement prevents credential theft. Rather than permanent keys, management layers utilize \textit{Just-in-Time (JIT) Credentials}—short-lived, scoped tokens injected directly into execution environments for task durations only. This minimizes confused deputy attack windows, automatically revoking access rights when tasks complete \cite{website:cormack2025securingmcp}.
\end{itemize}

\paragraph{Conclusion}
Resource vetting bridges discovery and execution. Layering static provenance (code source verification), dynamic reputation (behavior verification), and data sovereignty (usage rights verification) establishes trusted collaboration contexts. Resources passing these vetting gates enter operational environments (Section~\ref{op-env}) for secure, mediated execution.

\subsection{Ecosystem Governance and Oversight}
In closed MAS, developers own logs and control agents. Open Agent Internets lack centralized visibility. Safe interoperability requires shared \textit{Ecosystem Monitoring} infrastructure identifying, tracking, and attributing actions across organizational boundaries \cite{bengio2025singaporeconsensus}.

\subsubsection{Identity and Attribution}
While vetting (Section~\ref{vetting}) verifies gate credentials, governance requires session-long attribution. Agent-web service interactions need standardized authentication protocols binding all API calls and transactions to verifiable identities (DID-signed headers). This ensures accountability for ultimate initiating entities even in long delegation chains \cite{bengio2025singaporeconsensus, south2025authenticateddelegationauthorizedai}.

\subsubsection{Provenance and Watermarking}
Managing agent output downstream impacts requires content provenance. Watermarking (text/media) and metadata tagging trace AI-generated content to source agents. Model provenance mechanisms help researchers track underlying model lineage, enabling rapid identification and recall of harmful versions propagating through networks \cite{bengio2025singaporeconsensus}.

\subsubsection{Shared Accountability Infrastructure}
Establishing culpability for harmful events (financial losses from cascading agent failures) in decentralized networks presents challenges. The Singapore Consensus proposes \textit{Interoperable Logging Infrastructure}. Like aviation flight recorders, high-stakes agent interactions need shared \textit{Incident Reporting Standards} capturing decision sequences without exposing proprietary logic. This enables macro-level \textit{Usage Tracing} detecting systemic anomalies—coordinated botnet attacks—invisible to individual agents \cite{bengio2025singaporeconsensus, chan2025infrastructureaiagents}.

\paragraph{Conclusion}
Governance transforms open ecosystem theoretical risks into manageable operational challenges. While protocols and vetting provide preventive security, ecosystem oversight enables necessary retroactive recourse. Provenance enforcement and interoperable audit trail maintenance ensure Agentic Internet transparency and accountability, preventing responsibility diffusion in decentralized automation.

\subsection{Lessons Learned \& Key Takeaways}
Building the Agentic Internet requires architecting distributed trust systems beyond API connections. Robust IMAS demands holistic infrastructure: standardized interoperability stacks enabling communication, dynamic discovery mechanisms locating peers, rigorous resource vetting establishing confidence, and pervasive governance ensuring accountability. These pillars resolve tensions between autonomous agent generative power and interconnected world security mandates, laying foundations for collaborative AI global economies.

% --- SECTION 7: CONCLUSION ---
% --- SECTION 7: CONCLUSION ---
\section{Conclusion and Research Roadmap} \label{conclusion}
This paper deconstructs the agentic AI security landscape across three architectural tiers. Our analysis reveals several cross-cutting principles: (1) capability expansion inherently creates attack surface growth; (2) safety must be architecturally co-designed rather than added post-development; and (3) securing isolated components remains insufficient without addressing emergent systemic risks. 

From our comprehensive analysis, we synthesize essential research directions mapped directly to architectural vulnerabilities identified in preceding sections:

\subsection{Single-Agent Security Foundations (Section~\ref{singleagent})}
Our deconstruction of single agents revealed critical unresolved challenges:

\paragraph{Cognitive Security:} Model-level vulnerabilities discussed in Section~\ref{singleagent} demonstrate urgent need for architectures resistant to indirect prompt injection \cite{greshake2023youvesignedforcompromising} and universal adversarial attacks \cite{zou2023universaltransferableadversarialattacks}. Research must advance beyond probabilistic defenses toward verifiable guarantees through: (1) formal verification techniques for LLM prompts; (2) neuro-symbolic architectures grounding LLM reasoning in verifiable logic; and (3) system vector encoding approaches embedding safety constraints directly in activation space rather than prompt text \cite{cao2025cantstealnothingmitigating}.

\paragraph{Memory Integrity:} Memory vulnerabilities examined in Section~\ref{singleagent} require cryptographically verifiable memory architectures addressing both poisoning \cite{dong2025practicalmemoryinjectionattack} and action hijacking \cite{zhang2025actionhijackinglargelanguage}. Critical research paths include: (1) immutable audit logs with zero-knowledge proofs for memory operations; (2) content attestation protocols preventing adversarial rewrites; and (3) provable deletion algorithms supporting GDPR compliance.

\paragraph{Tool Trust:} The exploitable tool invocation patterns identified in Section~\ref{singleagent} demand standardized tool security protocols. Priority challenges include: (1) enhanced metadata safeguards preventing toolflow hijacking \cite{shi2025promptinjectionattacktool}; (2) privilege scope containment through mediated execution \cite{website:cormack2025securingmcp}; and (3) rigorous formal evaluation frameworks quantifying tool compromise impact.

\subsection{Multi-Agent System Resilience (Section~\ref{mas})}
Our analysis of collaborative systems identified architectural vulnerabilities requiring systemic solutions:

\paragraph{Topology Security:} Control flow vulnerabilities detailed in Section~\ref{mas} establish urgent need for adversarially robust topologies. Research must address: (1) automatic verification of multi-agent topology integrity against adversarial perturbations \cite{zugner2018adversarial}; (2) consensus-resilient pattern libraries resistant to goal drift \cite{arike2025technicalreportevaluatinggoal}; and (3) adaptive control monitoring proving topology-level safety invariants against deliberate subversion.

\paragraph{Communication Security:} Inter-agent messaging vulnerabilities identified in Section~\ref{mas} require cryptographically secure protocols detecting and preventing adversarial collaboration. Priority challenges include: (1) zero-knowledge semantic attestation validating message integrity; (2) semantic steganalysis detecting covert agent collusion \cite{motwani2025secretcollusionaiagents}; and (3) quantifiable information flow control preventing privilege escalation through message passing.

\paragraph{Environment Hardening:} The operational environment risks analyzed in Section~\ref{mas} reveal critical challenges in maintaining system integrity. Research paths include: (1) formally verified environmental containment; (2) adversarially robust reward models resistant to Goodhart's Law exploitation \cite{amodei2016concreteproblemsaisafety}; and (3) rigorous simulation-reality gap analysis preventing misgeneralization \cite{shah2022goalmisgeneralizationcorrectspecifications}.

\subsection{Interoperable Ecosystem Governance (Section~\ref{imas})}
Our examination of open agent networks established critical infrastructure requirements:

\paragraph{Protocol Security:} The interoperability stack vulnerabilities revealed in Section~\ref{imas} necessitate cryptographically secure interaction methods. Priority research includes: (1) post-quantum security frameworks for agent communication; (2) protocol downgrade attack mitigation for meta-protocol negotiation; and (3) formal verification of security properties across changing protocol versions.

\paragraph{Discovery and Trust:} Trust establishment challenges explored in Section~\ref{imas} demand zero-knowledge discovery frameworks. Research must advance: (1) Sybil-resistant decentralized reputation systems \cite{muscariello2025agntcyagentdirectory, raskar2025beyonddns}; (2) cryptographic capability attestation without compromising privacy; and (3) automatic verification of capability claims through secure challenge-response protocols \cite{aminiranjbar2024dawndesigningdistributedagents}.

\paragraph{Ecosystem Accountability:} Governance foundations discussed in Section~\ref{imas} require decentralized monitoring infrastructure. Critical research directions include: (1) privacy-preserving logging systems for decision provenance; (2) cryptographically secure watermarking resistant to adversarial removal; and (3) formal verification of interoperable governance protocols.

\subsection{Cross-Cutting Research Paradigms}
Beyond tier-specific challenges, we identify methodology shifts required across the agentic security landscape:

\paragraph{Quantifiable Assurance:} Agentic security demands transition from qualitative safety claims to quantitative assurance metrics with formal guarantees. Research must develop standardized benchmarking frameworks providing mathematical bounds on safety properties, enabling objective comparison between architectural approaches.

\paragraph{Adversarial Co-Evolution:} Security analysis requires continuous red-team evolution matching growing agentic capabilities. This necessitates developing competitive automated adversarial testing environments capturing novel attack vectors before deployment.

\paragraph{Human-AI Control Interfaces:} Human oversight ultimately forms last-line defense against systemic failure. Critical research must advance human-centric security frameworks accounting for automation bias \cite{greenblatt2024aicontrolimprovingsafety} and cognitive limitations in monitoring complex systems.

The Internet of Agents represents unprecedented automation potential, but only through rigorous implementation of this research roadmap can we balance capability with appropriate security constraints. By systematically addressing the architectural vulnerabilities across all tiers, we can build not merely powerful agents, but a trustworthy ecosystem ensuring safety, accountability, and human control as these systems evolve.

% --- REFERENCES ---
\bibliographystyle{unsrtnat} 
\bibliography{ref.bib}

\end{document}